\documentclass[prd,12pt,aps,superscriptaddress,showpacs,tightenlines,floatfix,nofootinbib,amssymb]{revtex4}

\usepackage{latexsym}
\usepackage{graphicx}

\def\simge{
    \mathrel{\rlap{\raise 0.511ex
        \hbox{$>$}}{\lower 0.511ex \hbox{$\sim$}}}}
\def\simle{
    \mathrel{\rlap{\raise 0.511ex
        \hbox{$<$}}{\lower 0.511ex \hbox{$\sim$}}}}
\def\beqn{\begin{equation}}
\def\eeqn{\end{equation}}
\def\barr{\begin{eqnarray}}
\def\earr{\end{eqnarray}}
\def\bc{\begin{center}}
\def\ec{\end{center}}

\begin{document}

\title{
Excited Baryon Spectroscopy from Lattice QCD:\\
Finite Size Effect and Hyperfine Mass Splitting
}

\author{Kiyoshi Sasaki}
 \affiliation{Department of Physics, University of Tokyo, Tokyo 113-0033, Japan}
 
\author{Shoichi Sasaki}
 \affiliation{Department of Physics, University of Tokyo, Tokyo 113-0033, Japan}
 \affiliation{RIKEN BNL Research Center, Brookhaven National Laboratory,
             Upton, NY 11973}

\date{\today}

\begin{abstract}
A study of the finite-size effect is carried out for spectra of 
both ground-state and excited-state baryons in quenched lattice
QCD using Wilson fermions. Our simulations are performed at
$\beta=6/g^2 =6.2$ with three different spatial sizes, $La\sim$
1.6, 2.2 and 3.2 fm. It is found that the physical lattice size larger than 
3 fm is required for $\Delta$ states in all spin-parity 
($J^{P}=1/2^{\pm}, 3/2^{\pm}$) channels and also negative-parity 
nucleon ($N^*$) state ($J^{P}=1/2^{-}$) even in the rather heavy quark mass 
region ($M_{\pi}\sim1.0$ GeV). We also find a peculiar behavior
of the finite-size effect on the hyperfine mass splittings 
between the nucleon and the $\Delta$ in both parity channels.

\end{abstract}

\maketitle


%
%
\section{Introduction}
\label{sec:intro}

In the past several years, many lattice studies 
on excited baryon spectroscopy have been made 
within the quenched approximation after the 
pioneering attempts~\cite{NstarPioneers}. 
Although these calculations gave results of negative-parity baryon masses 
in gross agreement 
with experiments~\cite{{Sasaki:2001nf},{Gockeler:2001db},{Melnitchouk:2002eg},
{Zanotti:2003fx},{Mathur:2003zf},{Nemoto:2003ft},{Brommel:2003jm},{Burch:2004he},
{Guadagnoli:2004wm}}, the higher precision study requires 
accurate controls of systematic errors, which may arise
from the quenched approximation, chiral extrapolation, 
nonzero lattice spacing and finite-size effect.
The first and second errors might appear most
seriously in spectra of the excited states and should be 
entangled with each other due to possible decay processes.
This complicated issue, however, ought to be postponed
since the current dynamical simulations are still limited even
for the ground-state baryons. It is worth emphasizing that
the quenched lattice QCD is still useful to investigate
several long standing puzzles in excited baryon spectroscopy,
such as the level ordering problem between the negative-parity nucleon
$N^*(1535)$ and the Roper resonance $N'(1440)$,
the structure of the $\Lambda(1405)$ and so on~\cite{Leinweber:2004it}.

Instead of using small lattice spacing, many calculations
have adopted sophisticated lattice discretization schemes for fermions,
the improved fermion actions~\cite{{Gockeler:2001db},{Melnitchouk:2002eg},
{Zanotti:2003fx},{Nemoto:2003ft}} or
chiral fermions~\cite{{Sasaki:2001nf},{Mathur:2003zf},{Brommel:2003jm},{Burch:2004he}},
in excited baryon spectroscopy to reduce the cutoff effect 
and improve the chiral behavior.
However, there is not so much attention paid to the finite-size effect.
Indeed, most of all simulations are performed with spatial lattice sizes
$La \sim 1.6-2.2$ fm. This is because the finite-size study
has been hardly done due to large computational costs. 
In this study, we utilize the standard Wilson fermion action,
the computational requirements of which are relatively cheap. 
Therefore, we systematically perform the finite-size study
for excited baryon spectroscopy.

For the finite-size effect, 
after the pioneering analytic consideration~\cite{Luscher:1985dn}, 
where the finite-size behavior of hadron masses obeys the exponential
 form $M_{\infty}-M_{L}\simeq
L^{-1}\exp(-L/L_0)$, 
it has been found that 
this behavior is rather described 
by a power law $\simeq L^{-n}$ with $n=2-3$~\cite{Fukugita:1992jj}. 
This power-law behavior can be interpreted 
by the ``wave function" squeezed on a finite lattice 
from the phenomenological point of view.
The dependence of the finite-size effect on the spatial boundary
condition which is found earlier~\cite{{Gupta:1982hu},{Martinelli:1982bm}}, 
can also be explained by this interpretation~\cite{Aoki:1993gi}. 
Ref.~\cite{Aoki:1993gi} reported that the lattice size
$La> 2.5$ fm is required for the nucleon ground-state ($N$) in quenched
lattice QCD.

In the non-relativistic quark model with a harmonics-oscillator
potential $H=\frac{1}{2m}\hat{p}^2+\frac{1}{2}m\omega^2\hat{r}^2$, 
the energy level and the root mean square radius can be estimated as 
$E_N=\hbar\omega (N+\frac{3}{2})$ and 
$r_{\rm rms} =\sqrt{\frac{\hbar}{m\omega}(N+\frac{3}{2})}$, respectively. 
The negative-parity nucleon ($N^*$) is identified as $N=1$ state, which corresponds 
to the first orbital excitation. It turns out that a ratio of the root mean square radii
between the $N^*$ and the $N$ states is estimated as 
$r_{\rm rms}^{N^*}/r_{\rm rms}^N=\sqrt{5/3}\sim1.3$. 
This means that the finite-size effect stemming from the hadron size squeezed is 
expected to become much serious for the excited-state baryons rather than the  ground-state
baryon. According to a crude estimate with this simple model, the lattice
size $La > 2.5 \times 1.3\sim3.2$ fm might be required for the excited baryon spectroscopy.

The organization of our paper is as follows. In Sec.\ref{sec:analytic}, we first review
the basic idea of the parity projection in lattice simulations. 
In recent years, the parity projection technique is greatly appreciated in determining 
the parity of the pentaquark state in lattice QCD~\cite{{Sasaki:2004vz},{Sasaki:2003gi}}.
Unfortunately, however, there are some initial confusions about the parity 
assignment~\cite{Sasaki:2004vz}. 
Thus, we describe details of the precise parity projection, which is 
always adopted by one of authors in excited baryon spectroscopy~\cite{{Sasaki:2001nf},
{Sasaki:2003gi},{Sasaki:2003xc}}.
For the spectroscopy of the spin-3/2 state, {\it i.e.} the $\Delta$ baryon, appropriate
spin projections are also described.
Sec.~\ref{sec:numerical} gives details of our Monte Carlo simulations and
results of the finite-size study for various baryon states, such as
the nucleon states in both parity channel ($J^{P}=1/2^{\pm}$) and
the $\Delta$ states in all spin-parity channels ($J^{P}=1/2^{\pm}, 3/2^{\pm}$).
Then, we discuss about mass splittings between 
hyperfine partners in both parity channels.
Finally, in Sec.~\ref{sec:conclusions} we summarize the present work.

%
%
\section{General Analytic Framework}
\label{sec:analytic}


\subsection{Parity projection}
\label{sec:ParityProj}

In general, the local baryon operator should couple to both parity states.
The local baryon operator is defined by the trilinear composite operator,
which is composed of a local diquark operator and 
a spectator-like quark field~\cite{Ioffe:1981kw} as
%
%
\beqn
{\cal O}_B(x)=\varepsilon_{abc}
\left(q_{a, i}^T(x)C\Gamma q_{b, j}(x)
\right) \Gamma' q_{c, k}(x)
\eeqn
where $\Gamma$ and $\Gamma'$ stand for
possible 16 Dirac matrices and $C$ is the charge conjugation matrix.
The superscript $T$ denotes transpose. Indices $a b c$ and $i j k$ have 
meanings as color and flavor respectively.
The intrinsic parity of this operator is defined by the parity transformation 
of internal quark fields ${\cal P}q({\vec x}, t){\cal P}^{\dag}=+\gamma_4 q(-{\vec x}, t)$ as
%
%
\beqn
{\cal P}{\cal O}_{B}^{(\eta)}({\vec x}, t){\cal P}^{\dag}
=
\eta \gamma_4 {\cal O}_{B}^{(\eta)}(-{\vec x}, t),
\eeqn
where $\eta$ denotes the intrinsic parity of ${\cal O}_{B}^{(\eta)}$. However, due to the simple relation
between the positive- and  negative-parity
operators: ${\cal O}_{B}^{(+)}(x)=\gamma_5 {\cal O}_B^{(-)}(x)$
{\it for the local baryon operator}, 
the resulting two-point correlation
functions are also related with 
each other~\cite{Sasaki:2001nf} as
%
%
\beqn
\langle 0| {\cal O}^{(+)}_{B}(x)\overline{\cal O}_B^{(+)}(y)|0\rangle
=-\gamma_5
\langle 0| {\cal O}^{(-)}_{B}(x)\overline{\cal O}_B^{(-)}(y)|0\rangle
\gamma_5\;.
\eeqn
This means that the two-point correlation function constructed
from the local baryon operator has overlap with both parity states.
Next, we consider the spin-1/2 baryon case as a typical example
for the precise parity projection. The zero-momentum two-point function 
is given by the sum over all spatial coordinates ${\vec x}$,
%
%
\beqn
G^{(\eta)}_{B}(t)=\sum_{\vec x}
\langle 0|
T\{
 {\cal O}^{(\eta)}_{B}({\vec x},t)\overline{\cal O}_B^{(\eta)}({\vec 0},0)
 \}
 |0\rangle \;.
\eeqn
This two-point correlation has the asymptotic form~\cite{Fucito:1982ip}
%
%
\beqn
G^{(\eta)}_{B}(t)=\frac{A_{\eta}}{2}
\left(1+{\rm sgn}(t)\gamma_4\right)e^{-M_{\eta}|t|}
-\frac{A_{-\eta}}{2}
\left(1-{\rm sgn}(t)\gamma_4\right)e^{-M_{-\eta}|t|}
\label{eq:Spin-1/2Func}
\eeqn
at large Euclidean time $t$. Here $M_{\eta}$ and $M_{-\eta}$ 
denote masses of the lowest-lying state in each parity channel.
The amplitude $A_{\eta}$ ($A_{-\eta}$) is
defined as $\langle 0| {\cal O}_{B}^{(\eta)}|B,\eta\rangle
=\sqrt{A_{\eta}}u^{(\eta)}_{B}$. Here, it is worth noting that
{\it anti-particle contribution of the opposite parity state is 
propagating forward in time}. Thus, the $+/-$ parity eigenstates
in the forward propagating contribution can be obtained by
choosing the appropriate projection 
$P^{(\eta)}_{\pm}=(1\pm \eta \gamma_4)/2$, which is given in 
reference to the intrinsic parity of the utilized operator, $\eta$.
This procedure is accomplished by taking the trace of the correlator
over spinor with the relevant projection operator as 
$\frac{1}{4}{\rm Tr}\{P^{(\eta)}_{\pm}G_{B}^{(\eta)}(t)\}$
where the factor of 1/4 is our choice of normalization.

The lattice simulation is, however, performed on a lattice
with finite extent $T$ in the time direction and (anti-)periodic
boundary conditions. The definition of the forward propagating 
contribution is slightly ambiguous since the backward propagating
can wrap the whole time range around the time boundary.
Hereafter, for a simplicity, we take $\eta=+$. Under the (anti-)periodic
conditions, the resulting correlation functions may be given by
%
%
\barr
G^{\rm p.b.c.}_{B}(t)&=&\sum_{n=-\infty}^{\infty} G_B(t+nT), \\
G^{\rm a.p.b.c.}_{B}(t)&=&\sum_{n=-\infty}^{\infty}(-)^{n} G_B(t+nT),
\earr
which certainly satisfy $G^{\rm p.b.c.}_{B}(t)=G^{\rm p.b.c.}_{B}(t+T)$
and $G^{\rm a.p.b.c.}_{B}(t)=-G^{\rm p.b.c.}_{B}(t+T)$ respectively.

For the region $0\le t < T$, those correlation functions are expressed
in the following forms:
%
%
\barr
G^{\rm p.b.c.}_{B}(t)
&=&
\sum_{n=0}^{\infty}A_{+}e^{-nM_{+}T}
\left[
\frac{1+\gamma_4}{2} e^{-M_{+}t}+\frac{1-\gamma_4}{2}e^{-M_{+}(T-t)}
\right] \nonumber \\
&&-
\sum_{n=0}^{\infty}A_{-}e^{-nM_{-}T}
\left[
\frac{1-\gamma_4}{2} e^{-M_{-}t}+\frac{1+\gamma_4}{2}e^{-M_{-}(T-t)}
\right],  \\ 
G^{\rm a.p.b.c.}_{B}(t)
&=&
\sum_{n=0}^{\infty}A_{+}e^{-nM_{+}T}
\left[
\frac{1+\gamma_4}{2} e^{-M_{+}t}-\frac{1-\gamma_4}{2}e^{-M_{+}(T-t)}
\right] \nonumber \\
&&-
\sum_{n=0}^{\infty}A_{-}e^{-nM_{-}T}
\left[
\frac{1-\gamma_4}{2} e^{-M_{-}t}-\frac{1+\gamma_4}{2}e^{-M_{-}(T-t)}
\right] \;.
\earr
For details of above calculation, see Appendix~\ref{appendix}.
In the large $T$ limit ($M_{+}T\gg1$ and $M_{-}T\gg 1$),
all $n\neq0$ terms are negligible in a summation with respect to $n$,
because of the factor $e^{-nM_{\pm}T}$. Thus, it turns out to be the form
%
%
\barr
G^{\rm p.b.c./a.p.b.c.}_{B}(t)
&\approx&
A_{+}
\left[
\frac{1+\gamma_4}{2} e^{-M_{+}t}\pm\frac{1-\gamma_4}{2}e^{-M_{+}(T-t)}
\right] \nonumber \\
&& -
A_{-}
\left[
\frac{1-\gamma_4}{2} e^{-M_{-}t}\pm\frac{1+\gamma_4}{2}e^{-M_{-}(T-t)}
\right]\;,
\earr
which has only a contribution from {\it the first wrap-round effect}.  
Clearly, even after the ``parity projection" as
$\frac{1}{4}{\rm Tr}\{P_{\pm}G_{B}^{\rm p.b.c./a.p.b.c.}(t)\}$,
there remain unwanted contaminations from the backward propagating contributions
due to the first wrap-round effect. However, one can easily find that
the linear combination of two correlation functions with periodic and
anti-periodic boundary conditions may prevent the first wrap-round effect:
%
%
\beqn
\overline{G}_{B}(t)=\frac{1}{2}\{
G_{B}^{\rm p.b.c.}(t)+G_{B}^{\rm a.p.b.c.}(t)
\}
\approx\left(\frac{1+\gamma_4}{2} \right)
A_{+}e^{-M_{+}t}-
\left(\frac{1-\gamma_4}{2}\right)A_{-} e^{-M_{-}t}.
\eeqn
Therefore, the parity projection are really accomplished through 
the operation $\frac{1}{4}{\rm Tr}\{P_{\pm}\overline{G}_{B}(t)\}$.

To achieve this in an efficient manner,
we adopt a linear combination of the quark propagators with
periodic and anti-periodic boundary conditions to
construct the baryon two-point correlation function~\cite{Sasaki:2001nf}.
The linear combination in the quark level automatically realizes
the linear combination in the hadronic level,
as shown below.
First, the baryon two-point correlation function $G_B(t)$ can be
schematically written by
%
%
\beqn
G_{B}(t)=\sum_{\vec x}{\rm tr}_{c}\{
S(x,0)\cdot S(x,0)\cdot S(x,0)
\}
\label{eq:ThreeQuark2Baryon}
\eeqn
with the quark propagator $S(x,0)$
where $x=({\vec x}, t)$ and $0=({\vec 0},0)$.
${\rm tr}_c$ denotes a trace over color indices.
We define the averaged quark propagator
%
%
\beqn
\overline{S}(x,0)=\frac{1}{2}\left(
S_{\rm P}(x,0)+S_{\rm AP}(x,0)
\right),
\eeqn
where $S_{\rm P}(x,0)$ and $S_{\rm AP}(x,0)$
are subject to the periodic and anti-periodic boundary conditions 
in the time direction such as
$S_{\rm P}({\vec x}, t; 0)=S_{\rm P}({\vec x}, t+T; 0)$ and  
$S_{\rm AP}({\vec x}, t; 0)=-S_{\rm AP}({\vec x}, t+T; 0)$.
By inserting this linear combination 
into Eq.(\ref{eq:ThreeQuark2Baryon}), the baryon two-point function
is written as 
%
%
\barr
G_{B}(t)&=&\frac{1}{8}\sum_{\vec x}{\rm tr}_{c}
\left\{
\left(
S_{\rm P}(x,0)+S_{\rm AP}(x,0)
\right)
\cdot
\left(
S_{\rm P}(x,0)+S_{\rm AP}(x,0)
\right)
\cdot
\left(
S_{\rm P}(x,0)+S_{\rm AP}(x,0)
\right)
\right\} \nonumber \\
&=&
\frac{1}{2}\left\{
G_{B}^{\rm p.b.c.}(t) + G_{B}^{\rm a.p.b.c.}(t)
\right\},
\earr
where
%
%
\barr
G_{B}^{\rm p.b.c.}(t)&=&
\frac{1}{4}\sum_{\vec x}
{\rm tr}_{c}
\left\{
S_{\rm P}(x,0)\cdot S_{\rm P}(x,0)\cdot S_{\rm P}(x,0)
+S_{\rm P}(x,0)\cdot S_{\rm AP}(x,0)\cdot S_{\rm AP}(x,0)\right. \nonumber \\
&&\left.
+S_{\rm AP}(x,0)\cdot S_{\rm P}(x,0)\cdot S_{\rm AP}(x,0)
+S_{\rm AP}(x,0)\cdot S_{\rm AP}(x,0)\cdot S_{\rm P}(x,0)
\right\}
\earr
and
%
%
\barr
G_{B}^{\rm a.p.b.c.}(t)&=&
\frac{1}{4}\sum_{\vec x}
{\rm tr}_{c}
\left\{
S_{\rm AP}(x,0)\cdot S_{\rm P}(x,0)\cdot S_{\rm P}(x,0)
+S_{\rm P}(x,0)\cdot S_{\rm AP}(x,0)\cdot S_{\rm P}(x,0)\right. \nonumber \\
&&\left.
+S_{\rm P}(x,0)\cdot S_{\rm P}(x,0)\cdot S_{\rm AP}(x,0)
+S_{\rm AP}(x,0)\cdot S_{\rm AP}(x,0)\cdot S_{\rm AP}(x,0)
\right\}.
\earr
$G_{B}^{\rm p.b.c.}(t)$ has even numbers of $S_{\rm AP}(x,0)$ and
then are satisfied with $G_{B}^{\rm p.b.c.}(t)=G_{B}^{\rm p.b.c.}(t+T)$.
On the other hand, $G_{B}^{\rm a.p.b.c.}(t)$ includes odd numbers of
$S_{\rm AP}(x,0)$ and is totally subject to the anti-periodic boundary
condition, $G_{B}^{\rm a.p.b.c.}(t)=-G_{B}^{\rm a.p.b.c.}(t+T)$.
We confirm that a linear combination of the quark propagators with
periodic and anti-periodic boundary conditions makes a cancellation
for the primal reflection from the time boundary in the baryon two-point function.
This is also true for the meson two-point function as same as the baryon case.

\subsection{Spin projection}
\label{sec:SpinProj}

In this subsection, we describe the spin projection, which is essential
to deal with spin-3/2 baryon, namely the $\Delta$ state. 
We choose a interpolating operator for the $\Delta$,
more specifically $\Delta^{++}$, as a simple trilinear composite
operator:
%
%
\beqn
{\cal O}_{\mu}^{\Delta}(x)=\epsilon_{abc}\left(u_a^T(x)C\gamma_\mu u_b(x)\right)u_c(x), 
\eeqn
where $u(x)$ represents the up quark field.
This operator has the structure of the Rarita-Schwinger spinor
(``vector-spinor")
so that the two-point function constructed from this operator
can couple to both spin-3/2 and spin-1/2 states~\cite{Leinweber:1992hy}.
We consider the two-point function at zero spatial momentum, which is 
given by
%
%
\beqn
G_{\mu \nu}^{\Delta}(t)=\sum_{\vec x}\langle 0|
T\{{\cal O}_{\mu}^{\Delta}({\vec x},t)\overline{\cal O}_{\nu}^{\Delta}
({\vec 0},0)\}
|0\rangle .
\label{eq:Spin-3/2Func}
\eeqn
Note that above $\Delta$ correlator can couple to the spin-3/2 state
only if neither of $\mu, \nu$ are temporal indices. For the spatial Lorentz indices, 
$i,j=1,2,3$, Eq. (\ref{eq:Spin-3/2Func}) is expressed 
by the orthogonal sum of spin-3/2 and spin-1/2 components:
%
%
\begin{equation}
  G^{\Delta}_{ij}(t)=
  \left(\delta_{ij}-\frac{1}{3}\gamma_i\gamma_j\right)
  G^{\Delta}_{3/2}(t)
  +\frac{1}{3}\gamma_i\gamma_j
  G^{\Delta}_{1/2}(t),
\end{equation}
with appropriate spin projection operators~\cite{Benmerrouche:1989uc}.
Respective spin parts, $G^\Delta_{3/2}(t)$ and $G^\Delta_{1/2}(t)$, 
possess both positive- and negative-parity contributions
as same as Eq.({\ref{eq:Spin-1/2Func}).
Some of excited $\Delta$ states such as the spin-1/2 $\Delta$ states 
can be accessed by the correlator $G^{\Delta}_{1/2}(t)$.
These spin projected correlators are given by 
%
%
\barr
G^{\Delta}_{3/2}(t)&=&\frac{3}{2}
G^{\Delta}_{ii}(t)-\frac{1}{2}\sum_{k}\gamma_{i}\gamma_{k}
G^{\Delta}_{ki}(t), \\
G^{\Delta}_{1/2}(t)&=&\sum_{k}
\gamma_{i}\gamma_{k}G^{\Delta}_{ki}(t),
\earr
where an index $k$ should be summed over all spatial directions,
but any specific choice is available for an index $i$. In this paper, we calculate all 
three direction for the spatial index $i$ and take an average of them to get
the possible reduction of statistical errors.

According to the previous subsection, we perform the appropriate parity projection
to apply to $G^{\Delta}_{3/2}(t)$ and $G^{\Delta}_{1/2}(t)$ as well. Then, we finally access
to the four different spin-parity states of the
$\Delta$ baryon, which correspond to the spin-3/2 positive-parity state
($\Delta_{3/2}$), the spin-3/2 negative-parity state ($\Delta^{*}_{3/2}$),
the spin-1/2 positive-parity state ($\Delta_{1/2}$)
and the spin-1/2 negative-parity state ($\Delta^{*}_{1/2}$). Of course, 
the $\Delta_{3/2}$ is the ground state of the $\Delta$ baryon.
All others are the excited $\Delta$ baryons.

%
%
\section{Numerical Results}
\label{sec:numerical}


We generate ensembles of the quenched QCD 
configurations with the standard 
single-plaquette action at $\beta=6/g^2=6.2$
with three different lattice sizes, 
$L^3\times T=24^3 \times 48$, $32^3 \times 48$ 
and $48^3 \times 48$, 
and compute the quark propagators 
by using Wilson fermion action at several values of the
hopping parameter $\kappa$, which cover the range 
$M_{\pi}\approx 0.6 - 1.2$ GeV.
Details of simulations are summarized in Tables~\ref{tab:simulation_pamam}
and \ref{tab:gluonic_pamam}.

For the update algorithm, we utilize the Metropolis algorithm
with 20 hits at each link update. The first 10000 sweeps are discarded 
for thermalization. The ${\cal O}$(200-300) gauge ensembles in each 
simulation are separated by 
1200 ($L$=48), 800 ($L$=32) and 600 ($L$=24)
sweeps.
For the matrix inversion, we use BiCGStab algorithm~\cite{Frommer:1994vn} 
and adopt the convergence condition $|r|< 10^{-8}$ for  the residues.
We calculate a simple point-point quark propagator with a source location
at $t_{\rm src}=6$. To perform the precise parity projection, we adopt 
a procedure to take an average of two quark propagators which are 
subject to periodic and anti-periodic boundary conditions in time, as
described in Sec. \ref{sec:analytic}. We use the conventional interpolating
operators, $\varepsilon_{abc}(u^{T}_{a}C\gamma_5 d_b)u_c$ for the
nucleon and $\varepsilon_{abc}(u^{T}_{a}C\gamma_\mu u_b)u_c$
for the $\Delta$ respectively.

All calculations were done on a Hitachi SR8000 parallel computer 
at KEK - High Energy Accelerator Research Organization, 
using the extended code based on the Lattice Tool Kit (LTK)~\cite{Choe:2002pu}. 
In this paper, we analyze the baryon-mass spectra in either parity channels with 
the conventional single exponential fit.

\subsection{Spatial lattice-size dependence} 
\label{subsec:finitevolume}

\indent
We first discuss the finite-size effect on masses of all measured baryons.
It is important to investigate how large size of physical lattice is required to neglect the finite-size effect on exited baryon spectroscopy.
For this purpose, we perform numerical simulations at three 
different lattice sizes, $L^3\times T=24^3 \times 48$, $32^3 \times 48$ 
and $48^3 \times 48$ for $\kappa=0.1506$ and 0.1497.
Quenched $\beta=6.2$ corresponds to a lattice cutoff of $a^{-1}=2.913$ GeV, 
which is set by the Sommer scale~\cite{Guagnelli:1998ud}. 
Thus, the spatial extents in our study correspond to 
$La\simeq 1.6,~2.2$ and 3.2 fm in the physical unit.
We will calculate the finite-size corrections to the baryon mass on each finite volume 
by comparison with values evaluated in the infinite volume limit. 

We utilize the phenomenological powe- law formula 
to take the infinite volume limit for the observed masses as
%
%
\begin{equation}
aM_L = aM_\infty  + c L^{-n}.
\label{Eq:PowerLaw}
\end{equation}
This power-law behavior can be interpreted by the 
hadron ``wave function" squeezed on a finite lattice 
from the phenomenological point of view~\cite{Fukugita:1992jj}. 
As for the quenched simulation, Ref.\cite{Aoki:1993gi}
reported that the range of $n=1-2$ is preferable for
the ground state of the nucleon. In this paper, we use the three
different powers, $n=1,2$ and 3, and then determine
which power-law behavior is most favorable for fitting data.

In Figs.\ref{fig:InfVol.Each}-\ref{fig:InfVol.DELT1}
the lattice-size dependence of each baryon mass
($N$, $N^*$, $\Delta_{3/2}$, $\Delta^{*}_{3/2}$,
$\Delta_{1/2}$, $\Delta^{*}_{1/2}$) is shown for
two hopping parameters ($\kappa=0.1497$ and 0.1506), which
correspond to the relatively heavier quark masses. 
All data included in Figs. \ref{fig:InfVol.Each}-\ref{fig:InfVol.DELT1}
are summarized in Table \ref{tab:fitted_mass_vol}.
The quoted errors in figures represent only the statistical errors,
which are obtained by a single elimination jack-knife method.
Horizontal dashed lines represent the values in the infinite volume limit,
which are evaluated in the case of $n=2$, together with their one 
standard deviation. A summary of masses in the infinite volume limit,
which are guided by various power-law behaviors, is given in 
Table \ref{tab:infvol_mass}. In the case of the $N^*$ state, which receives the 
largest finite-size effect, the power two ($n=2$) is most preferable 
because of $\chi^2/N_{DF}\sim 1$.

For the nucleon case, 
we confirm that the finite-size effect is very small in the relatively
heavier quark mass region. As shown in Fig. \ref{fig:InfVol.Each},
all the data are located  within $1\sigma$ of the value 
in the infinite volume limit at $\kappa=$0.1497 and 0.1506.
We do not see any serious finite-size effect on the mass of 
the ground-state nucleon even at the smallest lattice size $La \simeq 1.6$ fm.
In the mass of the negative-parity nucleon, 
a serious finite-size effect is not seen in the range of 
the spatial lattice size from 1.6 fm to 2.2 fm. This feature is consistent with that 
reported in Ref.~\cite{Gockeler:2001db}, in spite of the
fact that their observed tendency of the size effect is opposite.
Remarked that almost all subsequent lattice calculations to study the 
negative-parity nucleon are performed at the lattice size less 
than 2.2 fm. However, we find that the mass of the negative-parity nucleon 
suffers from the large finite-size effect in the range of 
the spatial lattice size from 2.2 fm to 3.2 fm. This means that 
the finite-size effect on the mass of the negative-parity baryon is unexpectedly
severe and the spatial size is required at least 3 fm to remove this effect.
What is the origin of the finite-size effect on the mass spectrum?
In the phenomenological point of view, the ``wave function" of baryon 
should be squeezed due to the small volume~\cite{Fukugita:1992jj}. 
The kinetic energy of internal quarks inside baryon is supposed to increase 
and thus the total energy of the three-quark system, 
which corresponds to the mass of baryon, should be pushed up.  Such effect 
is expected to become serious for the excited state rather than the ground state. 
This intuitive picture seems to account for the observed trend of decreasing mass as
the lattice size increases.

In the spin-3/2 $\Delta$ channel, however, we find that the finite-size effect has a 
different pattern: the ground-state mass of the spin-3/2 $\Delta$ becomes large as the lattice 
size increases. This behavior may originate from a hyperfine interaction. 
This possibility will be discussed later.
On the other hand, the negative-parity state of the spin-3/2 $\Delta$, 
namely the $\Delta_{3/2}^*$, has the same pattern of the finite-size effect
observed in the $N^*$ spectrum, while the finite-size correction of the $\Delta_{3/2}^*$ 
is relatively smaller than that of the $N^*$. As for the spin-1/2 $\Delta$ channel, all data 
at different lattice sizes are roughly consistent with each other within relatively large errors. 
However, the significant finite-size effect may be hidden behind the large statistical errors.

The finite-size effects on the masses of each baryon are summarized 
in Table~\ref{tab:fin_vol_eff}.  It is noteworthy that the finite-size 
corrections of $\Delta$ states ($J^P=3/2^{\pm}$) and negative-parity nucleon $N^*$ 
at spatial size $La$=1.6 fm can be seen even in the heavy quark region 
where that of the nucleon is almost negligible. 
Moreover, as we can learn from Figs.\ref{fig:InfVol.Each}-\ref{fig:InfVol.DELT1}, 
the spatial lattice size is required to be as large as 3 fm
to remove the finite-size effect. 
The systematic error stemming from the finite-size effect 
at spatial size $La\simeq 3.2$ fm, 
$|(m_\mathrm{3.2\hspace{2pt}fm}-m_\infty)/m_\infty|$, 
is smaller than 2\% for all measured states. 
Therefore, in the later discussion, we analyze data obtained 
on the largest lattice size of $48^3 \times 48$ for spectroscopy
of all hadrons.

Finally, we would like to comment on other possible formula 
such as the exponential form:
%
%
\begin{equation}
aM_L = aM_\infty  + c L^{-1}\exp(-L/L_0),
\end{equation}
which is inspired by the L{\"u}scher's formula for the asymptotic finite-size dependence of
stable particle masses~\cite{Luscher:1985dn}. 
Here $1/L_0 \sim M_{\pi}$ can be expected in the phenomenological sense. 
The full three-parameter fits tend to yield considerably low values of $1/L_0$.
The obtained values are as low as $10^{-4}$ in lattice units. It is clearly inconsistent with
a relation $1/L_0 \sim M_{\pi}$ since our simulations are performed 
in the range of $M_\pi \sim 0.2-0.5$ in lattice units. If the parameter $L_0$ is fixed 
by $1/L_0 = M_{\pi}$, the resulting $\chi^2/N_{DF}$ from 
the two-parameter fits is no longer reasonable. Therefore, the finite-size behavior
that we observed here can be described by a power-law formula (\ref{Eq:PowerLaw})
rather than above exponential formula.

\subsection{Chiral extrapolation}
\label{subsec:chiral}

All data of  hadron masses computed on lattice with spatial size $La\simeq 3.2$ fm
are tabulated in Tables \ref{tab:fitted_mass1} and \ref{tab:fitted_mass2}.
We perform a covariant single exponential fit to two-point functions 
of each hadron in respective fitting ranges.
All fits have a confidence level larger than 0.15 and $\chi^2/N_{DF}<1.5$. 
Our adopted ranges for fitting are basically determined by plateaus of each effective mass.

Next, we extrapolate masses of all calculated hadrons to the chiral limit 
using two types of fitting formula:
\begin{equation}
  aM_H = c_0 + c_2(aM_\pi)^2
  \label{eqn:chiral_linear_fit}
\end{equation}
or
\begin{equation}
  (aM_H)^2 = d_0 + d_2(aM_\pi)^2,
  \label{eqn:chiral_curve_fit} 
\end{equation}
where $c_0$, $c_2$, $d_0$, $d_2$ are numerical constants. 
We evaluate the systematic error by the difference between
the chiral limit values obtained by two different fitting formulae.
Hereafter, Eq.~(\ref{eqn:chiral_linear_fit}) and (\ref{eqn:chiral_curve_fit}) 
are referred as the ``linear'' fit and the ``curve'' fit, respectively. 
The chiral limit values and their values of fitting $\chi^2$ are listed in Table~\ref{table:ChiralLimit}.
For the $\rho$ meson, the nucleon and the $\Delta_{3/2}$, in accordance with 
resulting $\chi^2$ values, we can determine that the curve fit is 
much preferable. However, for the other baryons, either fits 
yield acceptable values of $\chi^2$ because of the relatively large
statistical errors on fitted data. However, the systematic error stemming
from the difference of two fitting results is less than 10\%. We then adopt 
the curve fit for all hadrons as the final analysis of chiral extrapolation.

In Figs.~\ref{fig:NuclCh}-\ref{fig:Delt1Ch}, we show squared masses of positive-
and negative-parity baryons as a function of squared pion mass in each ($I$, $J$) channel.
Circle symbols correspond to the negative-parity state (solid circles) and the positive-parity
state (open circles). The extrapolated points in the chiral limit are represented by
star symbols. In Table~\ref{tab:data_exp}, respective values in the chiral limit
for all calculated baryons are listed and also compared with
experimental values.
Two types of input ($r_0$ input and $M_{\rho}$ input) 
are taken to expose the dependence on the choice of input to set a scale.
We should keep in mind that the systematic errors stemming from this dependence 
are around 5\%, the value of which exceeds the amount of the statistical errors in the case of 
the nucleon and the $\Delta_{3/2}$. Instead, we quote various mass ratios, which
do not suffer from such systematic uncertainties:
%
%
\[
\begin{array}{rclr}
M_{\Delta_{3/2}}/M_{N}&=&1.28(4)(9) &({\rm Expt. :}\;\sim1.31)\\ 
M_{N^*}/M_{N}&=&1.61(10)(16)                &({\rm Expt. :}\;\sim1.63) \\ 
M_{\Delta^*_{3/2}}/M_{N^*}&=&1.28(7)(7)& ({\rm Expt. :}\;\sim1.11)  \\
M_{\Delta^*_{3/2}}/M_{\Delta_{3/2}}&=&1.61(7)(13) &({\rm Expt. :}\;\sim1.38) 
\end{array}
\]
where the second quoted errors correspond to the systematic errors, which
are estimated from the difference in the central values obtained by two 
types of chiral extrapolation. The mass ratio between the nucleon and its 
parity partner (or its hyperfine partner)
shows a good agreement with the experimental values within statistical errors.
However, mass ratios that include
$M_{\Delta_{3/2}^{*}}$ are overestimated by about 15\%. 
Our calculations are performed using the relatively heavy quark masses
($M_{\pi}/M_{\rho}\sim 0.66-0.96$). The long chiral extrapolation
is performed so that the evaluated values should not be taken too seriously.
Indeed, both the ``linear" fit and the ``curve" fit do not include
a term linear in $aM_{\pi}$, which is responsible for the expected leading behavior 
close to the chiral limit in the quenched approximation.~\cite{Labrenz:1996jy}

Finally we stress that the level ordering in $\Delta$ spectra~\footnote[1]{
In our results, $\Delta^*_{3/2}$ and $\Delta^*_{1/2}$, both of
which belong to the same $SU(6)$ multiplet, are mostly degenerate
within statistical errors. Strictly speaking, we have not seen clear splitting between 
them. 
},
$M_{\Delta_{3/2}}<M_{\Delta^{*}_{1/2}}\simle M_{\Delta^{*}_{3/2}}<M_{\Delta_{1/2}}$,
is well reproduced in comparison to corresponding $\Delta$ states, which are
all ranked as four stars on the Particle Data Table~\cite{Eidelman:2004wy}.
In addition, it is worth mentioning that a signal of $\Delta(1750)$ ($I=3/2$ and 
$J^P=1/2^{+}$), which is the weakly established state (one star)~\cite{Eidelman:2004wy}, 
cannot be seen in our data.

\subsection{Hyperfine mass splitting}
\label{subsec:hyperfine}

\indent
In this subsection, 
we discuss our numerical results on 
the hyperfine mass splittings in both parity channels
({\it e.g. $M_{\Delta_{3/2}}-M_N$ and $M_{\Delta_{3/2}^*}-M_{N^*}$}). 
In the quark potential model, 
if the inter-quarks potential is central force 
and independent of the flavor and spin, 
the hamiltonian has spin-flavor $SU(6)$ symmetry~\cite{Capstick:2000qj}. 
Under the exact spin-flavor $SU(6)$ symmetry, 
masses of resonance in the same $SU(6)$ multiplet, 
{\it e.g.} ${\bf 56}$-plet for the nucleon and the $\Delta_{3/2}$ particle 
or ${\bf 70}$-plet for negative-parity $N^*$ and $\Delta_{3/2}^*$ states, 
should be degenerate. 
However, the actual baryon spectra show evident violations of this symmetry. 
In the non-relativistic quark models, violations of the spin-flavor $SU(6)$
symmetry are caused by spin-dependent interactions~\cite{Capstick:2000qj}.

First, we consider the finite-size effect on the hyperfine mass splittings 
 ($N-\Delta_{3/2}$ and $N^*-\Delta_{3/2}^*$).
The lattice-size dependences of the hyperfine mass splittings are shown
in Figs.~\ref{fig:InfVol.Hyp} and \ref{fig:InfVol.HypSTR}.
We observe an unique feature in both parity channels. Each hyperfine 
splitting becomes small as the spatial lattice size decreases~\footnote[2]{
The same feature is observed in the charmonium spectrum~\cite{Choe:2003wx}.
The hyperfine splitting between $\eta_c$ and $J/\psi$ diminishes as 
the spatial lattice size decreases if $La\simle 1.3$ fm.}.
This feature 
is rather peculiar from the viewpoint of the non-relativistic quark models.
The hyperfine interaction may be derived from one-gluon exchange 
as a spin-spin component of the Fermi-Breit type interaction.
Thus, the hyperfine mass splitting originates from a contact interaction between 
colored quarks~\cite{Capstick:2000qj}. 
As the size of a hadron decreases, probability of finding two quarks at the same 
spatial point inside hadron increases, so that the hyperfine mass splitting would increase.
However, the observed finite-size effect where
the hyperfine mass splitting diminishes as the lattice size decreases
is opposite from above naive expectation.
At present, we do not have a simple interpretation of the finite-size 
behavior of the hyperfine mass splitting in our data of quenched lattice QCD.
This peculiar behavior may suggest some other origin 
of the hyperfine interaction~\cite{Capstick:2000qj} rather than one-gluon exchange.
Finally, it is worth mentioning that the observed size dependence of the hyperfine mass
splitting accounts for the opposite pattern of the finite-size effect on masses of 
the $\Delta_{3/2}$ and also our observation that the finite-size effect
for the $\Delta_{3/2}^{*}$ state is slightly milder than the $N^{*}$ state.

We shortly comment on the splittings between pairs of parity partner 
($N-N^*$ and $\Delta_{3/2}-\Delta_{3/2}^*$).
It is found that these splittings become small as the lattice size 
decreases in both $N$ and $\Delta$ channels. All data including
hyperfine mass splittings are tabulated in Table~\ref{tab:hyp_mass}.

Next, we discuss the quark-mass dependence of the hyperfine mass splitting.
In Fig.~\ref{fig:HyperFine}, the hyperfine mass splittings between the nucleon
and the $\Delta_{3/2}$ are plotted as a function of sum of 
the nucleon and  $\Delta_{3/2}$ masses by using data at the largest
spatial lattice size, $La\simeq 3.2$ fm. 
We find that the hyperfine mass splitting becomes large as the quark mass decreases.
This feature shows a good agreement with the general form of the spin-dependent interaction 
in the non-relativistic quark models where the spin-dependent interaction is proportional to
inverse powers of the quark mass~\cite{Capstick:2000qj}. Of course, one can easily find 
such the feature in the actual baryon spectra in the wide range
from the light (up, down) sector to the charm sector.
Fig.~\ref{fig:HyperFine} includes some experimental 
points (stars), which correspond to 
spin 1/2 and 3/2 doublets;  $N(939)-\Delta(1232)$, $\Sigma(1192)-\Sigma(1385)$, $\Xi(1315)-\Xi(1530)$ and $\Sigma_c(2455)-\Sigma_c(2520)$.
Our data including the chiral extrapolated value, $M_{\Delta_{3/2}}-M_{N}=0.262(11)$ GeV,  is fairly consistent with those experimental points. 
It is also found that the quark-mass dependence for the hyperfine mass splitting between the $\Delta_{3/2}^{*}$ and the $N^*$ is consistent with the naive expectation, which is deduced from the explicit mass dependence of the spin-dependent interaction in the non-relativistic quark models.

Finally, it is important to note that the value of  the 
hyperfine mass splitting should be sensible to the leading discretization 
errors of the Wilson fermion action, which may induce the extra
chromomagnetic-moment interaction at finite lattice spacing.
We however believe that qualitative features of the finite-size dependence and
the quark-mass dependence, which are observed in this paper, do not change
with respect to discretization ${\cal O}(a)$ errors. 

%
%
\section{Conclusions}
\label{sec:conclusions}

In this paper, we have studied the finite-size effect
on masses of various baryons: the nucleon states in both parity channels
($J^P=1/2^{\pm}$) and the $\Delta$ states in all spin-parity channels
($J^P=1/2^{\pm},3/2^{\pm}$). Our quenched lattice simulations were
employed  at relatively weaker coupling, $\beta = 6/g^2 = 6.2$, where 
the cut-off scale ($\sim 3$ GeV) is definitely higher than mass scale of all
observed baryons ($\sim 1-2$ GeV). Three different lattice sizes, 
$La\simeq 1.6, 2.2$ and 3.2 fm, were utilized to examine how large lattice size 
is required for excited baryon spectroscopy.

We have found the considerable finite-size effect on masses of 
all $\Delta$ states and negative-parity nucleon $N^*$ 
even in the relatively heavy quark region where the finite-size effect on the nucleon
is almost negligible. The finite-size behavior that we observed can be described
by a power law $M_{\infty}-M_{L}\propto L^{-n}$ with $n\approx 2$ rather than
the exponential form $\propto L^{-1}\exp(-L/L_0)$. This observation is consistent with
that reported in Ref~\cite{Fukugita:1992jj}. 
If the finite-size effect is kept as small as a few percent level, the spatial lattice size 
$La \simge 3$ fm should be required for excited baryons, especially for the negative-parity
nucleon.

The finite-size behavior of the power law might originate from the phenomenological
point of view as the squeezed hadron size. Indeed, we confirmed that the rather large 
finite-size effect appears for excited baryons, the ``wave function" of which ought 
to be extended rather than that of the ground state. The squeezed ``wave function" may
increase the total energy of the three-quark system because of a gain
in kinetic energy of internal quarks inside hadron so that the mass is expected to
be pushed up as the lattice size decreases. However, the $\Delta_{3/2}$ state, which
is the lowest lying state in the $\Delta$ channel, reveals the opposite pattern of the
size effect. The mass of  the $\Delta_{3/2}$ decreases as the lattice size decreases.
This peculiar behavior might attribute to the finite-size effect on the hyperfine interaction.

According to the $SU(6)$ quark model, the nucleon and the $\Delta_{3/2}$ belong 
to the same $SU(6)$ multiplet so that the difference of wave functions between
the $N$ and  the $\Delta_{3/2}$ is induced by the hyperfine interaction. 
Our study of the finite-size 
effect were performed in the relatively heavy quark mass region, where the finite-size 
effect on the nucleon mass is almost negligible. Therefore, the finite-size effects on the 
$\Delta_{3/2}$ can attribute to the effect of the hyperfine mass splitting.
Indeed, in the case of the $N^{*}-\Delta^{*}_{3/2}$
hyperfine mass splitting, we observed the same pattern of 
the finite-size effect where $N^{*}-\Delta^{*}_{3/2}$ splitting diminishes 
as the lattice spatial size decreases, while either $N^{*}$ or $\Delta^{*}_{3/2}$
states yield the expected size effect where both masses increase as the lattice size decreases.
However, the hyperfine interaction induced by one-gluon exchange, which is mainly
a contact interaction between colored quarks, may not  correctly account for 
this peculiar behavior of the finite-size effect. It seems that our observed
finite-size effect on the hyperfine mass splitting suggests some other origin of 
the hyperfine interaction for baryons rather than one-gluon exchange.

%

%
%


\section*{Acknowledgments}
It is a pleasure to acknowledge A. Nakamura and C. Nonaka for helping us 
develop codes for our lattice QCD simulations from their open
source codes (Lattice Tool Kit~\cite{Choe:2002pu}).
We also appreciate T. Hatsuda for helpful suggestions.
This work is supported by the Supercomputer Projects No.102 
 (FY2003) and No.110 (FY2004) of High Energy Accelerator Research Organization (KEK).
S.S. thanks for the support by JSPS Grand-in-Aid for 
Encouragement of Young Scientists (No.15740137).	      

%
%
\appendix

\section{(anti-)periodic time dependence for finite T}
\label{appendix}

\indent
Let us consider the functional form of the typical two-point correlation function 
$G(t)=e^{-M|t|}$ under the periodic (anti-periodic) boundary condition in time. 
The desired function, in the finite extent $T$ with the periodic
boundary condition, should be formed as
%
%
\beqn
G_{\rm p.b.c.}(t)=\sum_{n=-\infty}^{\infty}G(t+nT),
\eeqn
which certainly satisfies $G_{\rm p.b.c.}(t)=G_{\rm p.b.c.}(t+T)$.
Recall that the given correlation function $G(t)$ clearly has the
time-reflection symmetry $G(t)=G(-t)$\footnote[1]{
In the case of $G(t)={\rm sgn}(t)e^{-M|t|}$, a property $G(t)=-G(-t)$ should be
utilized instead of the time-reflection symmetry.}.
Thus, one can rewrite $G_{\rm p.b.c.}(t)$ for $0\le t<T$ as
%
%
\begin{eqnarray}
G_{\rm p.b.c.}(t)&=&\sum_{n=0}^{\infty}G(t+nT)+\sum_{n=1}^{\infty}G(nT-t) \nonumber \\
&=& \sum_{n=0}^{\infty} e^{-M(t+nT)} + \sum_{n=1}^{\infty} e^{-M(nT-t)}  \nonumber \\
&=&
\sum_{n=0}^{\infty} e^{-nMT} \left[
e^{-Mt}+e^{-M(T-t)} \right] .
\label{pbc} 
\end{eqnarray}
In the case of the anti-periodic boundary condition, where
$G_{\rm a.p.b.c.}(t)=-G_{\rm a.p.b.c.}(t+T)$, one obtains
the following general form:
%
%
\begin{eqnarray}
G_{\rm a.p.b.c.}(t)&=&
\sum_{n=-\infty}^{\infty}(-)^nG(t+nT) \nonumber \\
&=&
\sum_{n=0}^{\infty}(-)^nG(t+nT)+\sum_{n=1}^{\infty}(-)^nG(nT-t) \nonumber \\
&=&
\sum_{n=0}^{\infty} (-)^n e^{-nMT} \left[
e^{-Mt}-e^{-M(T-t)} \right] .
\label{apbc}
\end{eqnarray}

We would like to make the following few comments.
First, we can derive familiar forms from Eqs.(\ref{pbc}) and (\ref{apbc}) as
the asymptotic form in the large $T$ limit ($MT>>1$).
$$
G_{\rm p.b.c./a.p.b.c.}(t)\rightarrow 
e^{-Mt}\pm e^{-M(T-t)}
$$
which has only a contribution from {\it the first wrap-round effect}. The $(n-1)$-th wrap-round
effect should be suppressed with the factor $e^{-nMT}$. 
Then, the linear combination of $G_{\rm p.b.c.}(t)$ and $G_{\rm a.p.b.c.}(t)$
yields 
$$
\overline{G}(t)=\frac{1}{2}\left[
G_{\rm p.b.c.}(t)+ G_{\rm a.p.b.c.}(t)
\right]\approx e^{-Mt}
$$
where the first wrap-round effect is canceled.

Secondary, we will see that $\overline{G}(t)$ 
corresponds to a periodic function in the finite
extent $2T$ as follows. 
One sums up all wrap-round contributions in
Eqs.(\ref{pbc}) and (\ref{apbc}), and then obtains
the exact functional forms:
%
%
\barr
G_{\rm p.b.c.}(t)&=&
\frac{1}{1-e^{-MT}}\left[
e^{-Mt}+e^{-M(T-t)} \right]
=
\frac{\cosh[M(t-\frac{T}{2})]}{\sinh[M\frac{T}{2}]} \\
G_{\rm a.p.b.c.}(t)&=&
\frac{1}{1+e^{-MT}}\left[
e^{-Mt}-e^{-M(T-t)} \right] 
=
-\frac{\sinh[M(t-\frac{T}{2})]}{\cosh[M\frac{T}{2}]}
\earr
where the (anti-)periodic time dependence for finite $T$ is given
by a $\cosh (\sinh)$ function.
Therefore, $\overline{G}(t)$ is given by
%
%
\beqn
\overline{G}(t)=\frac{1}{2}\left(
\frac{\cosh[M(t-\frac{T}{2})]}{\sinh[M\frac{T}{2}]}-
\frac{\sinh[M(t-\frac{T}{2})]}{\cosh[M\frac{T}{2}]}
\right)
=
\frac{\cosh[M(t-T)]}{\sinh[MT]}
\eeqn
The final expression is nothing but $G_{\rm p.b.c.}(t)$ with the $2T$-periodicity.

%
%

\newpage

\begin{table}[htbp]
\caption{
Simulation parameters for each volume studied in this work.
}
\label{tab:simulation_pamam}
\begin{ruledtabular}
\begin{tabular}{lccccc}
\hline
$\beta$ & lattice size ($L^3 \times T$) & kappa values & statistics & $\sim La$ (fm)\\
\hline
6.2  & $24^3 \times 48$ &  \{0.1506, 0.1497\} & 320 &  1.6\\ 
  & $32^3 \times 48$ &  \{0.1506, 0.1497\} & 240 & 2.2\\ 
  & $48^3 \times 48$ &  \{0.1520, 0.1506, 0.1497, 0.1489, 0.1480\} & 210  &3.2\\
\hline
\end{tabular}
\end{ruledtabular}
\end{table}

\begin{table}[htbp]
\caption{
The physical scale set by the Sommer parameter $r_0=0.5$ fm.
The value of $r_{0}/a$ is taken from Ref.~\protect\cite{Guagnelli:1998ud}. 
}
\label{tab:gluonic_pamam}
\begin{ruledtabular}
\begin{tabular}{cccc}
\hline
\hline
$\beta$ & $r_0/a$ & $a^{-1}$ (GeV) & $a$ (fm) \\
\hline
6.2 & 7.380(3) & 2.913 & 0.06775 \\ 
\hline
\hline
\end{tabular}
\end{ruledtabular}
\end{table}

\begin{table}[htbp]
\caption{Finite-size effect on masses of all measured baryons at $\beta=6.2$}
\label{tab:fitted_mass_vol}
\begin{ruledtabular}
\begin{tabular}{lccccccc}
$L^3 \times T$ & $\kappa$ & $a M_{N}$ & $a M_{N^*}$& $a M_{\Delta_{3/2}}$
& $a M_{\Delta_{3/2}^*}$ & $a M_{\Delta_{1/2}}$ & $a M_{\Delta_{1/2}^*}$\\
\hline
$24^3 \times 48$ 
& 0.1506 &  0.594(5) & 0.829(19) & 0.624(7) & 0.872(16) & 0.993(24) & 0.777(28)\\
& 0.1497 &  0.667(4) & 0.891(15) & 0.693(5) & 0.928(13) & 1.055(22) & 0.864(22)\\
\hline
$32^3 \times 48$
& 0.1506 &  0.597(4) & 0.808(15) & 0.628(7) & 0.878(13) & 0.985(27) & 0.835(25)\\
& 0.1497 &  0.670(3) & 0.873(13) & 0.697(4) & 0.928(12) & 1.043(25) & 0.901(21)\\
\hline
$48^3 \times 48$
& 0.1506 &  0.594(4) & 0.762(15) & 0.637(5) & 0.837(14) & 0.964(26) &0.814(28)\\
& 0.1497 &  0.668(3) & 0.832(13) & 0.702(4) & 0.892(12) & 1.023(25) &0.869(23)\\
\hline
\end{tabular}
\end{ruledtabular}
\end{table}

\begin{table}[htbp]
\caption{Infinite volume limit of various baryon masses at $\beta=6.2$}
\label{tab:infvol_mass}
\begin{ruledtabular}
\begin{tabular}{cl|cc|cc|cc|cc}
$\kappa$ & type & $a M_{N}$ & $\chi^2/N_{DF}$ & $a M_{N^*}$ & $\chi^2/N_{DF}$ 
& $a M_{\Delta_{3/2}}$ & $\chi^2/N_{DF}$ 
& $a M_{\Delta_{3/2}^*}$ & $\chi^2/N_{DF}$  \\
\hline
0.1506 & $L^{-3}$ & 0.597(4) & 0.46 & 0.763(16) & 1.54 & 0.638(5) & 0.42 & 0.846(14) & 3.06\\
              & $L^{-2}$ & 0.595(5) & 0.47 & 0.747(20) & 0.94 & 0.641(7) & 0.26 & 0.836(18) & 2.56\\
              & $L^{-1}$     & 0.595(9) & 0.47 & 0.697(36) & 0.45 & 0.651(12)&0.12 & 0.808(31) & 2.01\\
\hline
0.1497 & $L^{-3}$ & 0.669(4) & 0.37 & 0.833(13) & 1.70 & 0.703(4) & 0.22 & 0.898(12) &  2.51\\
              & $L^{-2}$ & 0.669(5) & 0.40 & 0.819(17) & 1.06 & 0.705(6) & 0.11 & 0.889(16) & 
2.00\\
              & $L^{-1}$     & 0.669(8) & 0.41 & 0.775(30) & 0.52 & 0.712(10)&0.03 & 0.862(27) & 1.47\\
\hline
\end{tabular}
\end{ruledtabular}
\end{table}

\begin{table}[htbp]
\caption{Summary table of the finite-size effect 
  on various baryon masses for $\kappa=0.1506$.}
\label{tab:fin_vol_eff}
\begin{ruledtabular}
\begin{tabular}{cccc}
\hline
baryon     & as increase $L$ & 
$\sim(M_\mathrm{1.6\hspace{2pt}fm}-M_\infty)/M_\infty$ &
$\sim|M_\mathrm{3.2\hspace{2pt}fm}-M_\infty|/M_\infty$ \\
\hline
$N$                              & $\to$            & less than $1\%$ & less than $1\%$ \\
$N^*$                         & $\searrow$ &        $11\%$   &  $2\%$ \\
$\Delta_{3/2}$       & $\nearrow$ &        $-3\%$  & less than $1\%$ \\
$\Delta_{3/2}^*$  & $\searrow$  &        $4\%$    & less than $1\%$ \\
\hline
\end{tabular}
\end{ruledtabular}
\end{table}


\begin{table}[htbp]
\caption{Fitted masses of pion, $\rho$-meson and nucleon ($J^P=1/2^{\pm}$) states 
at $\beta=6.2$ and lattice volume $48^3\times 48$.}
\label{tab:fitted_mass1}
\begin{ruledtabular}
\begin{tabular}{c|cc|cc|cc|cc}
$\kappa$ & $a M_\pi$ & range & $a M_{\rho}$ & range & $a M_{N}$ & range & $a M_{N^*}$
& range \\
\hline
$\kappa_c$
                  & 0 & N/A       &0.248(3)&N/A& 0.332(9) & N/A & 0.534(33) & N/A\\
 0.1520  & 0.2141(8) & [15,30] &0.324(2)&[14,30]& 0.469(5) 
 & [20,28] & 0.640(22) & [13,20] \\
 0.1506  & 0.3167(9) & [15,30] &0.388(1)&[14,30]& 0.594(4) 
 & [20,28] & 0.762(15) & [13,20] \\
 0.1497  & 0.3726(9) & [15,30] &0.430(1)&[14,30]& 0.668(3) 
 & [20,28] & 0.832(13) & [13,20] \\
 0.1489  & 0.4188(9) & [15,30] &0.467(1)&[14,30]& 0.730(3) 
 & [20,28] & 0.891(12) & [13,20] \\
 0.1480  & 0.4678(9) & [15,30] &0.509(1)&[14,30]& 0.799(3) 
 & [20,28] & 0.957(11) & [13,20] \\
\hline
\end{tabular}
\end{ruledtabular}
\end{table}

\begin{table}[htbp]
\caption{Fitted masses of $\Delta$ baryons ($J^{P}=3/2^{\pm}, 1/2^{\pm}$)
at $\beta=6.2$ and lattice volume $48^3\times 48$.}
\label{tab:fitted_mass2}
\begin{ruledtabular}
\begin{tabular}{c|cc|cc|cc|cc}
$\kappa$ & $a M_{\Delta_{3/2}}$ & range & $a M_{\Delta_{3/2}^*}$
& range & $a M_{\Delta_{1/2}}$ & range & $a M_{\Delta_{1/2}^*}$ & range\\
\hline
 $\kappa_c$
                  & 0.424(11)& N/A       & 0.681(25) & N/A 
                  & 0.798(38)& N/A       & 0.640(53) & N/A \\
 0.1520  & 0.535(9)  & [22,28] & 0.770(21) & [12,20]
                  & 0.883(34)& [10,15] & 0.732(54)& [13,20]\\
 0.1506  & 0.637(5)  & [22,28] & 0.837(14) & [12,20]
                  & 0.964(26)& [10,15] & 0.814(28)& [13,20]\\
 0.1497  & 0.702(4)  & [22,28] & 0.892(12) & [12,20]
                  & 1.023(25)& [10,15] & 0.869(23)& [13,20]\\
 0.1489  & 0.759(4)  & [22,28] & 0.943(12) & [12,20]
                  & 1.076(25)& [10,15] & 0.921(20)& [13,20]\\
 0.1480  & 0.822(4)  & [22,28] & 1.001(11) & [12,20]
                  & 1.136(25)& [10,15] & 0.981(18)& [13,20]\\
\hline
\end{tabular}
\end{ruledtabular}
\end{table}

%
%
\begin{table}[htbp]
\caption{Results of the chiral extrapolated masses in lattice units for all measured hadrons
by two types of fitting formula, the linear and curve fits.}
\label{table:ChiralLimit}
\begin{ruledtabular}
\begin{tabular}{cc|ccccccc}
type &
 &
$aM_{\rho}$ &
$aM_N$ &
$aM_{N^*}$ &
$aM_{\Delta_{3/2}}$ &
$aM_{\Delta_{3/2}^*}$ &
$aM_{\Delta_{1/2}}$ &
$aM_{\Delta_{1/2}^*}$ \\
\hline
  linear & (mass)              &  0.283(2) & 0.403(7) & 0.582(26) & 0.477(8)  & 0.703(22)  & 0.818(35)
  & 0.670(44)\\
  & ($\chi^2/N_{DF}$)  & 6.20(1.67) & 9.62(3.37)           & 0.62(0.45)          & 1.57(0.99)        
   & 0.03(0.08)             & 0.01(0.02)    & 0.01(0.05)\\
\hline
 curve  & (mass)              & 0.248(3) & 0.332(9) & 0.534(33) & 0.424(11) & 0.681(25)& 0.798(38) 
 & 0.640(53) \\
  & ($\chi^2/N_{DF}$)  & 1.59(0.63) & 0.03(0.14)         & 0.07(0.12)              & 0.05(0.15)           
  & 0.17(0.21)          & 0.01(0.03)    & 0.01(0.05)\\
\end{tabular}
\end{ruledtabular}
\end{table}
%

\begin{table}[htbp]
\caption{
The third and forth columns list results of all measured baryon masses in GeV units, 
which are set by two different inputs, $r_0$ input and $M_{\rho}$ input.
The fifth and sixth list experimental values of the corresponding baryon and
its status in the Particle Data Table~\protect\cite{Eidelman:2004wy}.
The possible assignments of the $SU(6)\otimes O(3)$ supermultiplet are
also embedded into the final column.
}
\label{tab:data_exp}
\begin{ruledtabular}
\begin{tabular}{cc|ll|lc|c}
state &($I,J^P$) & our results  [GeV]  &  & physical baryon & (status) & $SU(6)\otimes O(3)$ \\
&& ($r_0$ input)& ($M_{\rho}$ input) & & &classification \\
\hline
$N$ & 
$(\frac{1}{2},\frac{1}{2}^{+})$ & 0.967(26) &  1.030(27) &$N(939)$ &**** & [{\bf 56},$0^+$]\\
$N^{*}$&
$(\frac{1}{2},\frac{1}{2}^{-})$  & 1.555(96) &  1.658(102)&$N(1535)$ $S_{11}$  &**** & [{\bf 70},$1^-$]\\
$\Delta_{3/2}$&
$(\frac{3}{2},\frac{3}{2}^{+})$ & 1.235(32) &  1.316(34)&$\Delta(1232)$ $P_{33}$ & **** & [{\bf 56,$0^+$]}\\
$\Delta^{*}_{3/2}$&
 $(\frac{3}{2},\frac{3}{2}^{-})$ & 1.983(72) &  2.114(77)&$\Delta(1700)$ $D_{33}$ &**** & [{\bf 70},$1^-$]\\
$\Delta_{1/2}$&
$(\frac{3}{2},\frac{1}{2}^{+})$ & 2.324(110) & 2.477(117)&$\Delta(1910)$ $P_{31}$ &**** & [{\bf 56},$2^+$] or [{\bf 70},$0^{+}$]\\
$\Delta^{*}_{1/2}$&
 $(\frac{3}{2},\frac{1}{2}^{-})$ & 1.864(154) & 1.987(164)&$\Delta(1620)$ $S_{31}$ &****&[{\bf 70},$1^-$]\\
\hline
\end{tabular}
\end{ruledtabular}
\end{table}

\begin{table}[htbp]
\caption{Finite-size effect on various mass splittings  at $\beta=6.2$}
\label{tab:hyp_mass}
\begin{ruledtabular}
\begin{tabular}{lcccccc}
$L^3 \times T$ &  $\kappa$ 
& $a M_{\Delta_{3/2}}$-$a M_{N}$ & $a M_{\Delta_{3/2}^*}$-$a M_{N^*}$  
& $a M_{N^*}$-$a M_{N}$ & $a M_{\Delta_{3/2}^*}$-$a M_{\Delta_{3/2}}$&
$a M_{\Delta_{1/2}}$-$a M_{\Delta_{1/2}^*}$\\
\hline
$24^3 \times 48$ 
& 0.1506 & 0.030(7) & 0.043(20) & 0.235(21) & 0.248(18)  &0.216(36) \\
& 0.1497 & 0.026(5) & 0.037(14) &  0.224(16) & 0.235(15) & 0.191(30) \\
\hline
$32^3 \times 48$
& 0.1506 & 0.030(6) & 0.070(17) &  0.211(17) & 0.250(15) & 0.150(37) \\
& 0.1497 & 0.027(4) & 0.055(13)  &  0.201(14) & 0.231(13) & 0.143(33) \\
\hline
$48^3 \times 48$
& 0.1506 & 0.043(5) & 0.075(16) &  0.168(18) & 0.200(16) & 0.151(35) \\
& 0.1497 & 0.035(4) & 0.060(12) &  0.164(15) & 0.190(14) & 0.154(31) \\
\hline
\end{tabular}
\end{ruledtabular}
\end{table}

%
%

\begin{figure}[htbp]
\begin{center}
\begin{minipage}[b]{5.25in}
\includegraphics[width=2.0in]{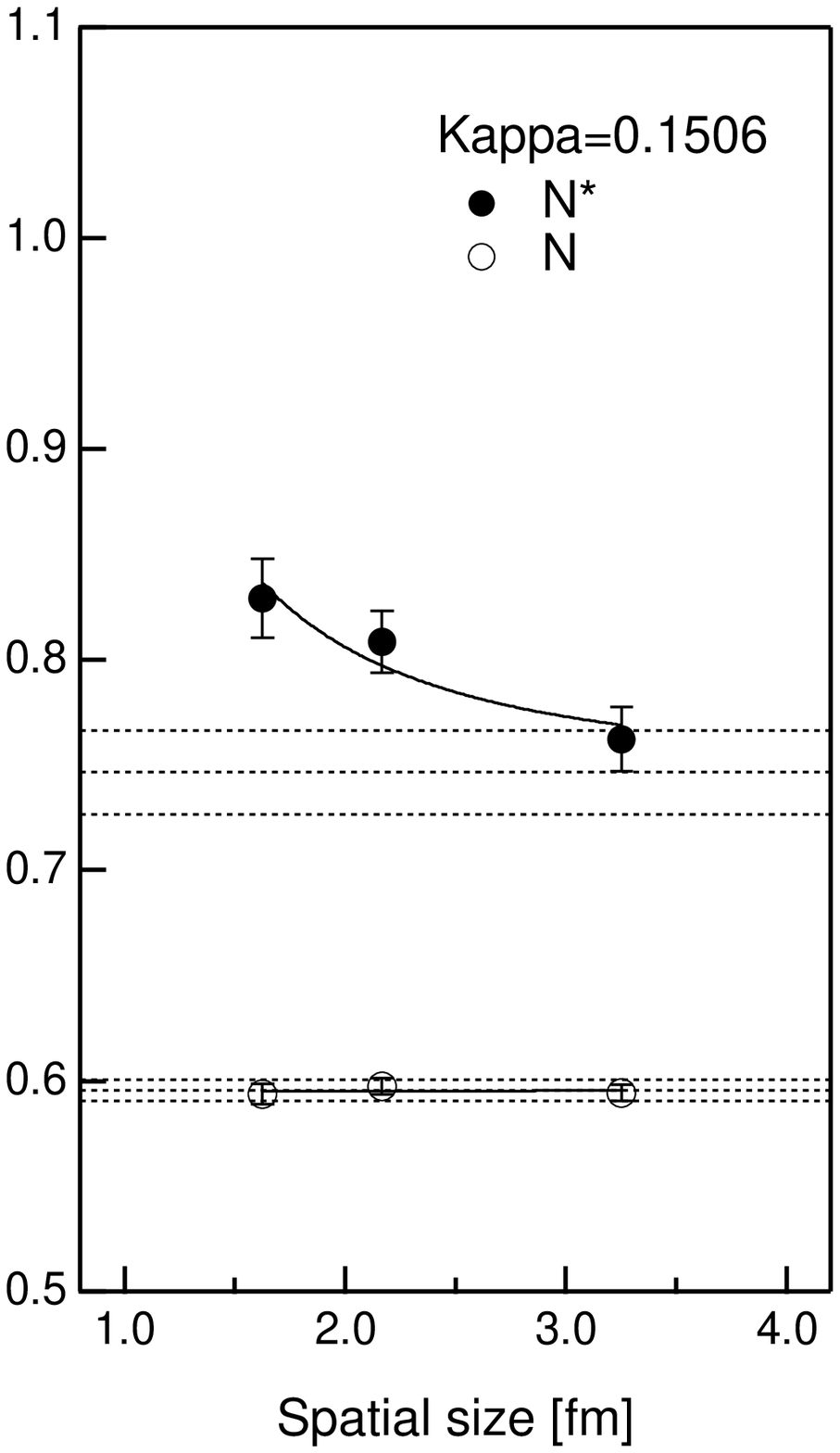}
\includegraphics[width=2.0in]{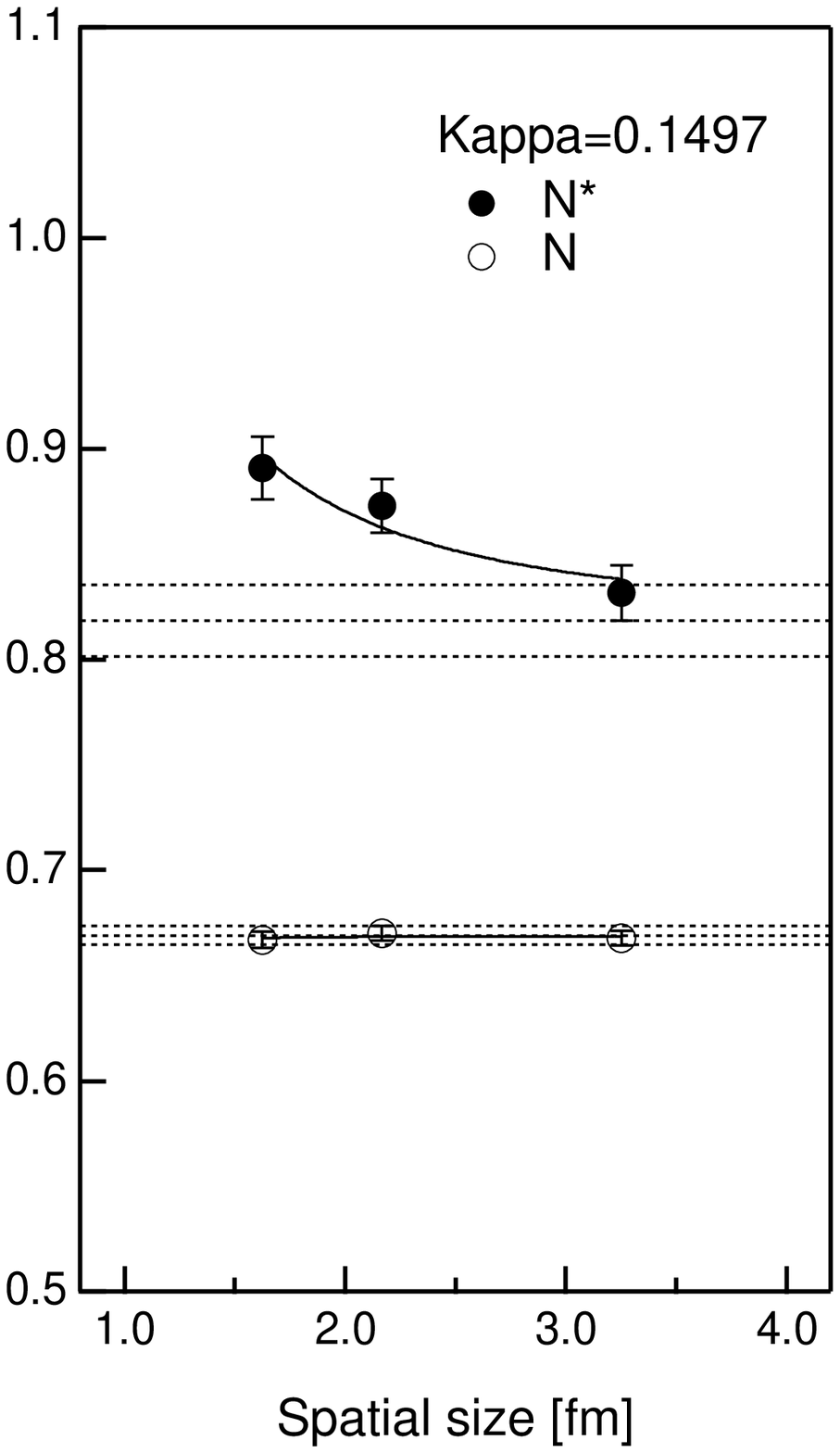}
\linebreak
\vfill
\end{minipage}
\end{center}
\caption{Masses of nucleons ($J^P=1/2^{\pm}$) in lattice
unit as functions of spatial lattice size in the physical unit for $\kappa=0.1506$
(left figure) and $\kappa=0.1497$ (right figure). Solid curves
are fits of the form $aM_L=aM_{\infty}+cL^{-n}$ with the value $n=2$.
Horizontal dashed lines represent extrapolated values for each parity state
in the infinite volume limit, together with their one standard deviation.
}
\label{fig:InfVol.Each}
\end{figure}

\begin{figure}[htbp]
\begin{center}
\begin{minipage}[b]{5.25in}
\includegraphics[width=2.0in]{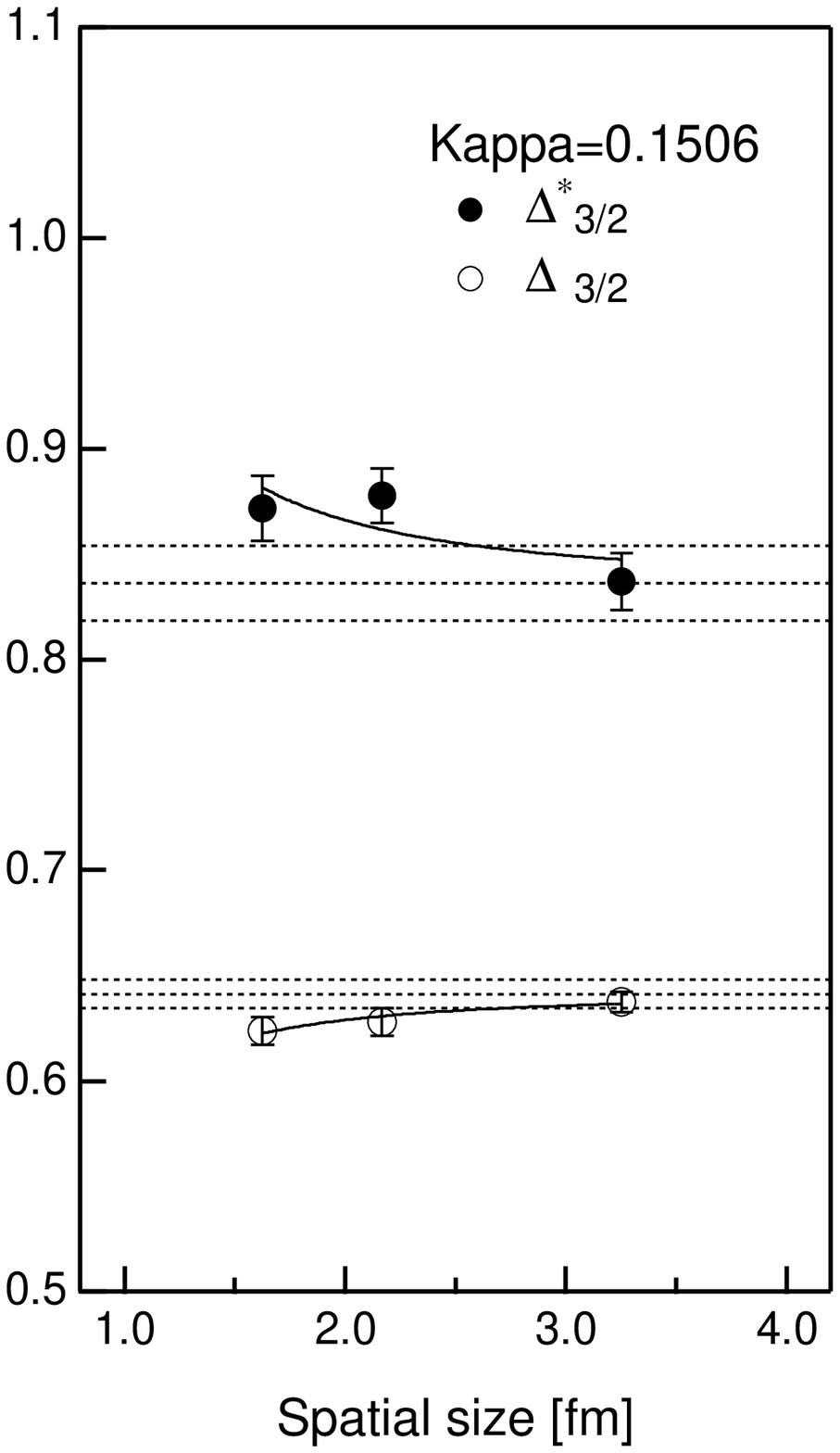}
\includegraphics[width=2.0in]{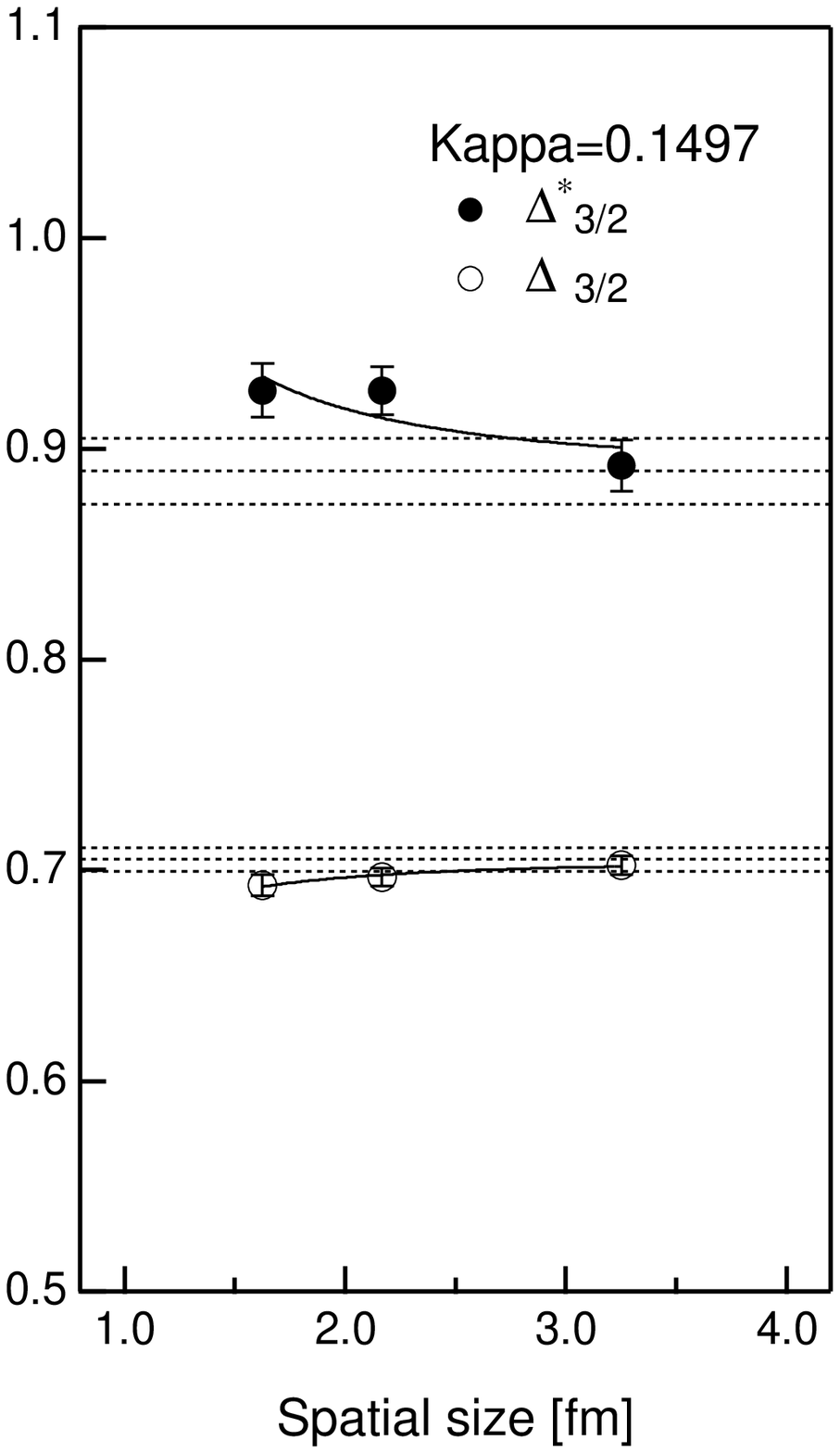}
\linebreak
\vfill
\end{minipage}
\end{center}
\caption{Masses of spin-3/2 $\Delta$ baryons in lattice
unit as functions of spatial lattice size in the physical unit for $\kappa=0.1506$
(left figure) and $\kappa=0.1497$ (right figure). 
Solid curves and dashed lines are defined as in 
Figure~\ref{fig:InfVol.Each}.
}
\label{fig:InfVol.DELT3}
\end{figure}

\begin{figure}[htbp]
\begin{center}
\begin{minipage}[b]{5.25in}
\includegraphics[width=2.0in]{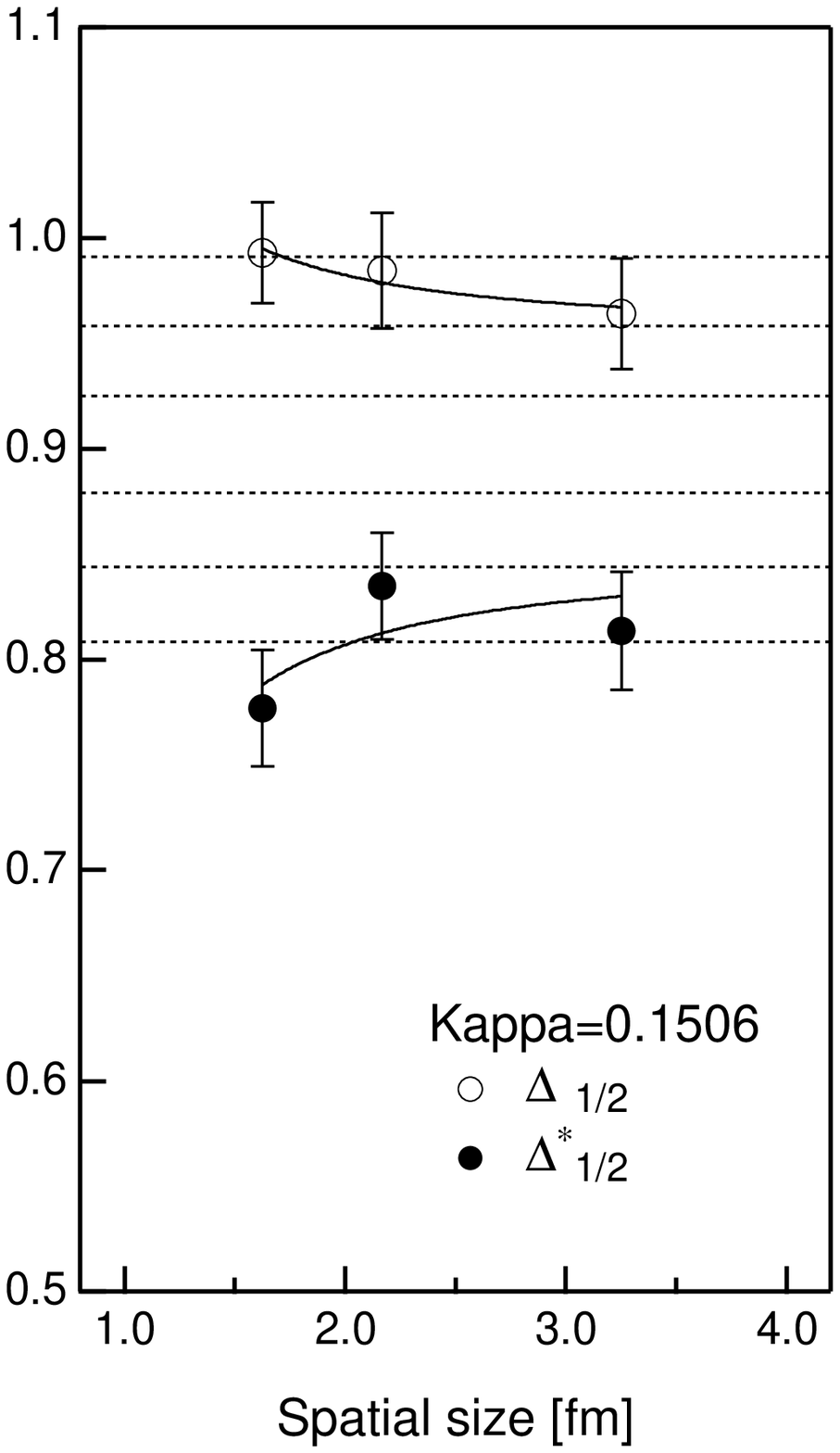}
\includegraphics[width=2.0in]{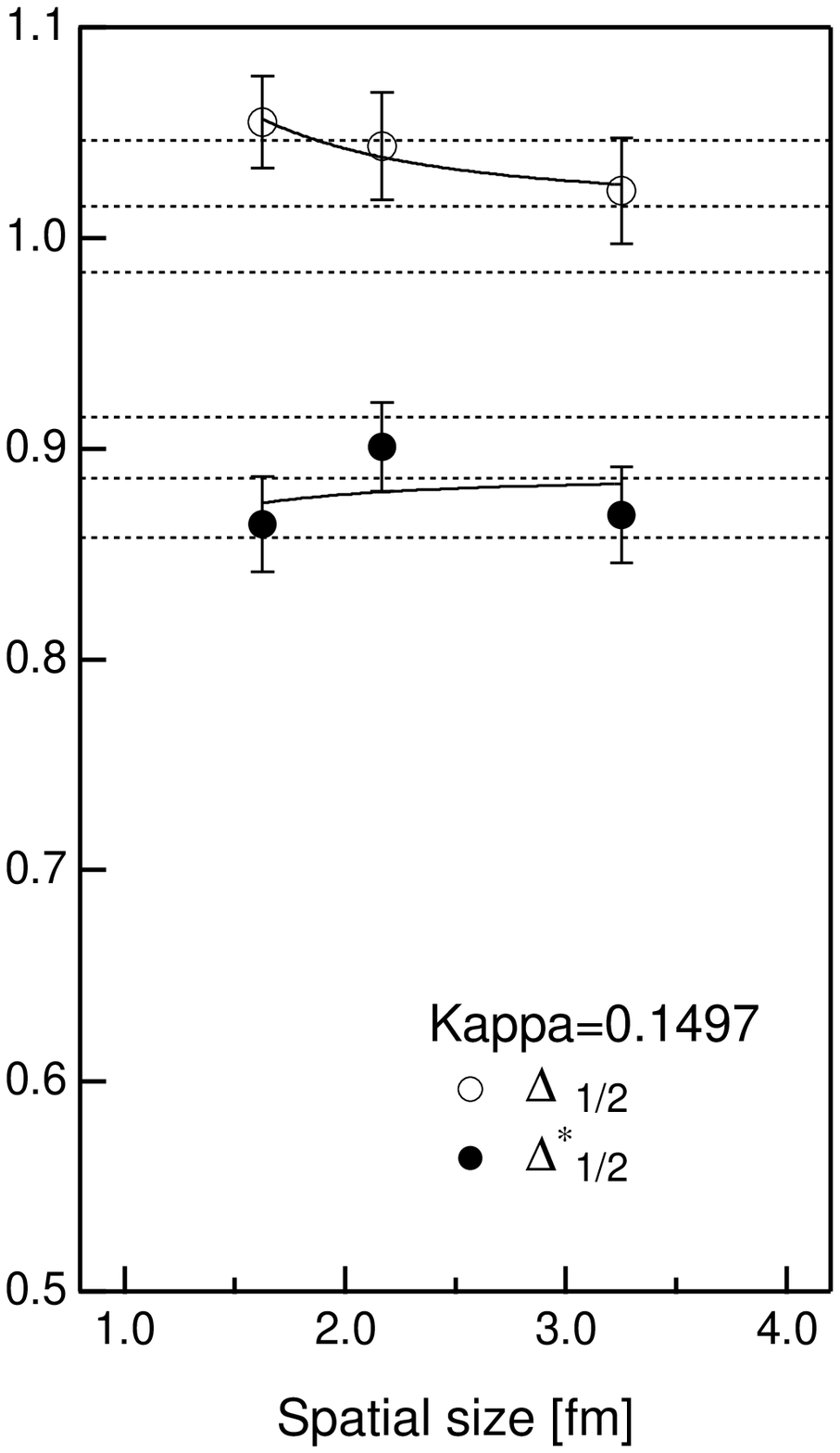}
\linebreak
\vfill
\end{minipage}
\end{center}
\caption{Masses of spin-1/2 $\Delta$ baryons in lattice
unit as functions of spatial lattice size in the physical unit for $\kappa=0.1506$
(left figure) and $\kappa=0.1497$ (right figure). 
Solid curves and dashed lines are defined as in 
Figure~\ref{fig:InfVol.Each}.
}
\label{fig:InfVol.DELT1}
\end{figure}

\begin{figure}[htbp]
\begin{center}
\includegraphics[width=3.0in]{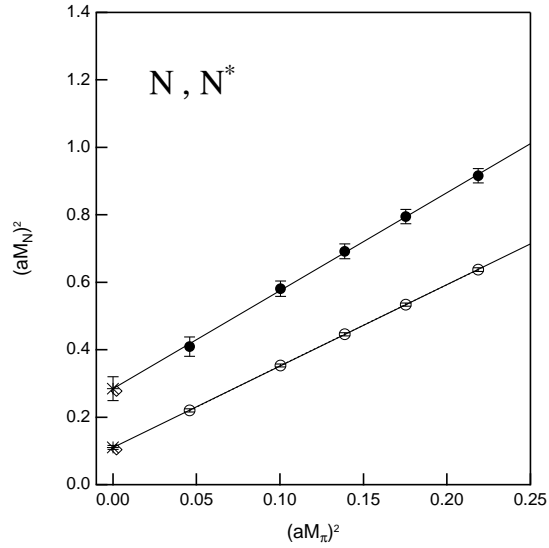}
\end{center}
\caption{Squared masses of nucleons ($J^P=1/2^{\pm}$) in lattice
unit as functions of squared pion mass.
Circle symbols correspond to the negative-parity state (solid circles) and
the positive-parity state (open circles). 
The extrapolated points in the chiral limit are represented by stars 
respectively. The experimental values are marked with lower and upper
open diamonds.}
\label{fig:NuclCh}
\end{figure}

\begin{figure}[htbp]
\begin{center}
\includegraphics[width=3.0in]{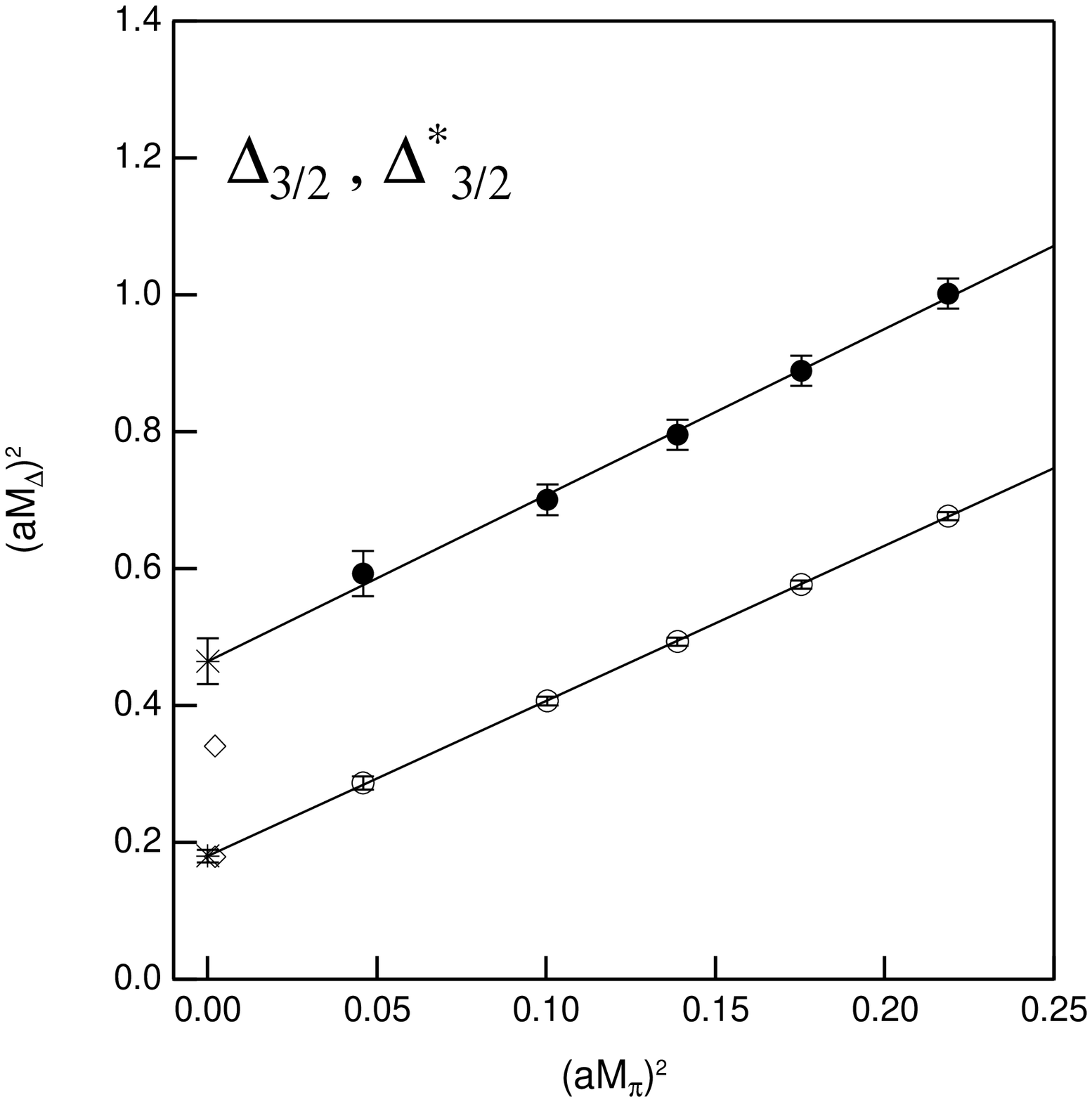}
\end{center}
\caption{Squared masses of $\Delta$ baryons ($J^P=3/2^{\pm}$) in lattice
unit as functions of squared pion mass.
All symbols are defined as in Figure \ref{fig:NuclCh}.
}
\label{fig:Delt3Ch}
\end{figure}

\begin{figure}[htbp]
\begin{center}
\includegraphics[width=3.0in]{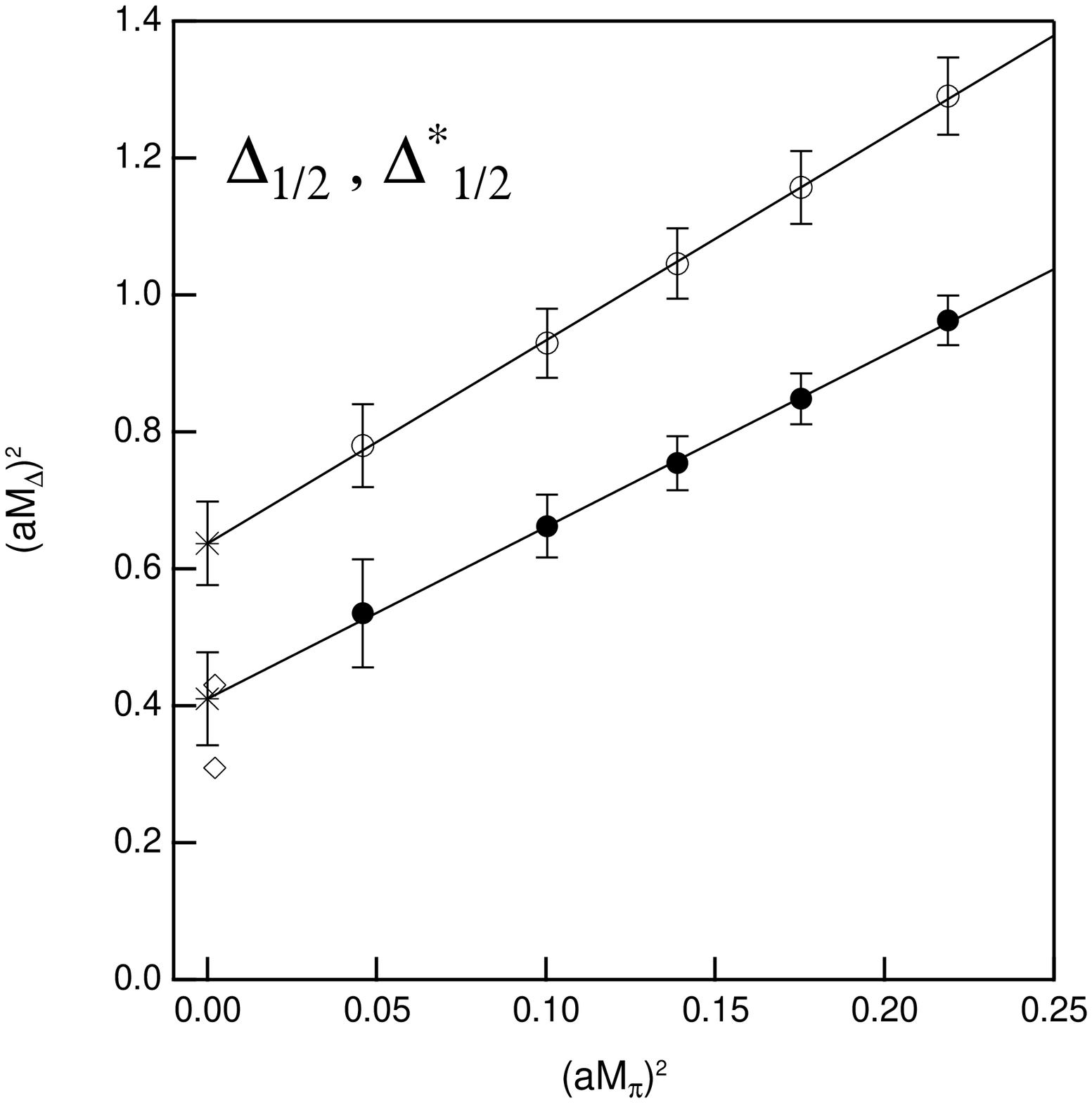}
\end{center}
\caption{Squared masses of $\Delta$ baryons ($J^P=1/2^{\pm}$) in lattice
unit as functions of squared pion mass.
All symbols are defined as in Figure \ref{fig:NuclCh}.}
\label{fig:Delt1Ch}
\end{figure}

\begin{figure}[htbp]
\begin{center}
\begin{minipage}[b]{5.25in}
\includegraphics[width=2.0in]{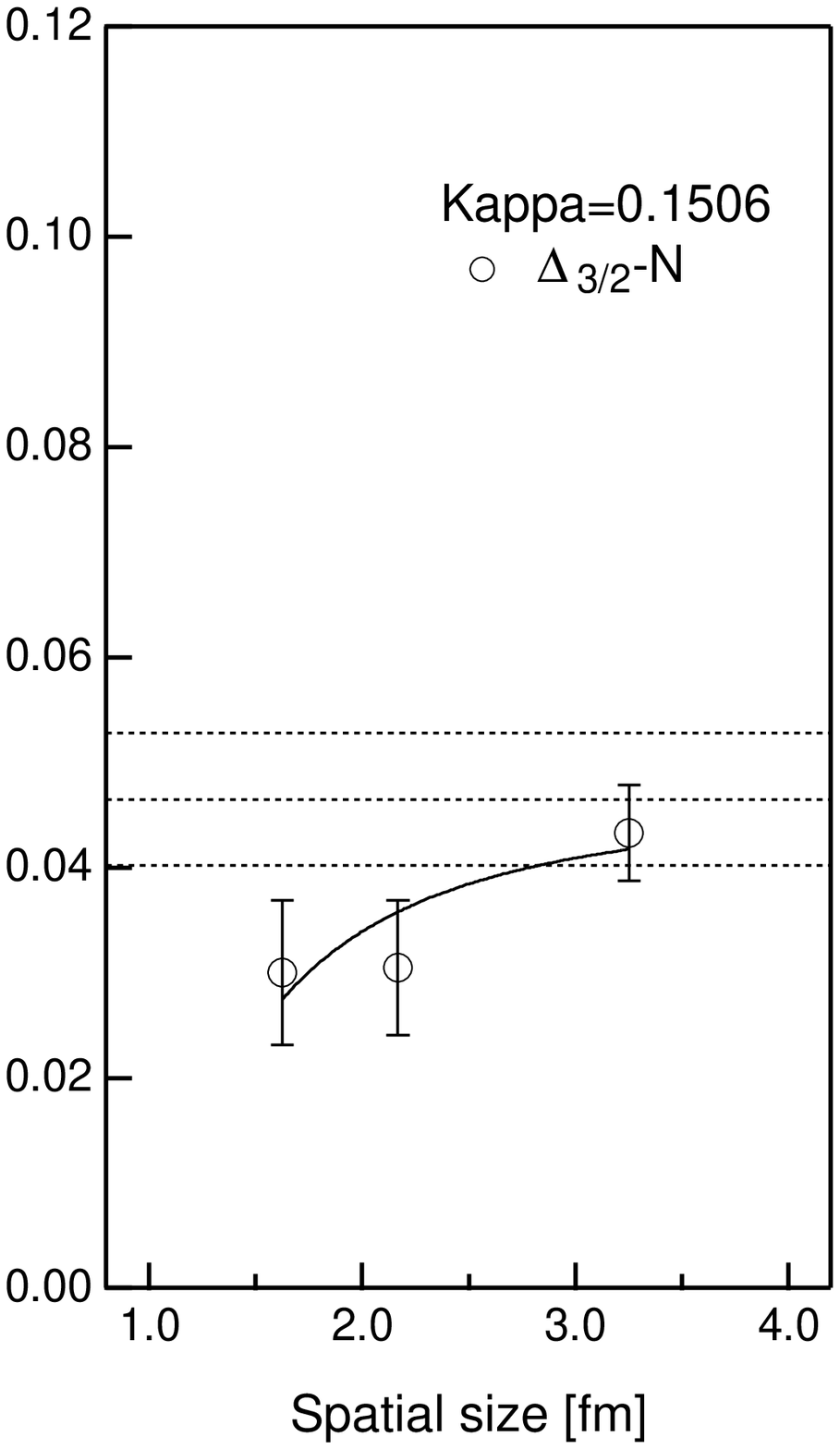}
\includegraphics[width=2.0in]{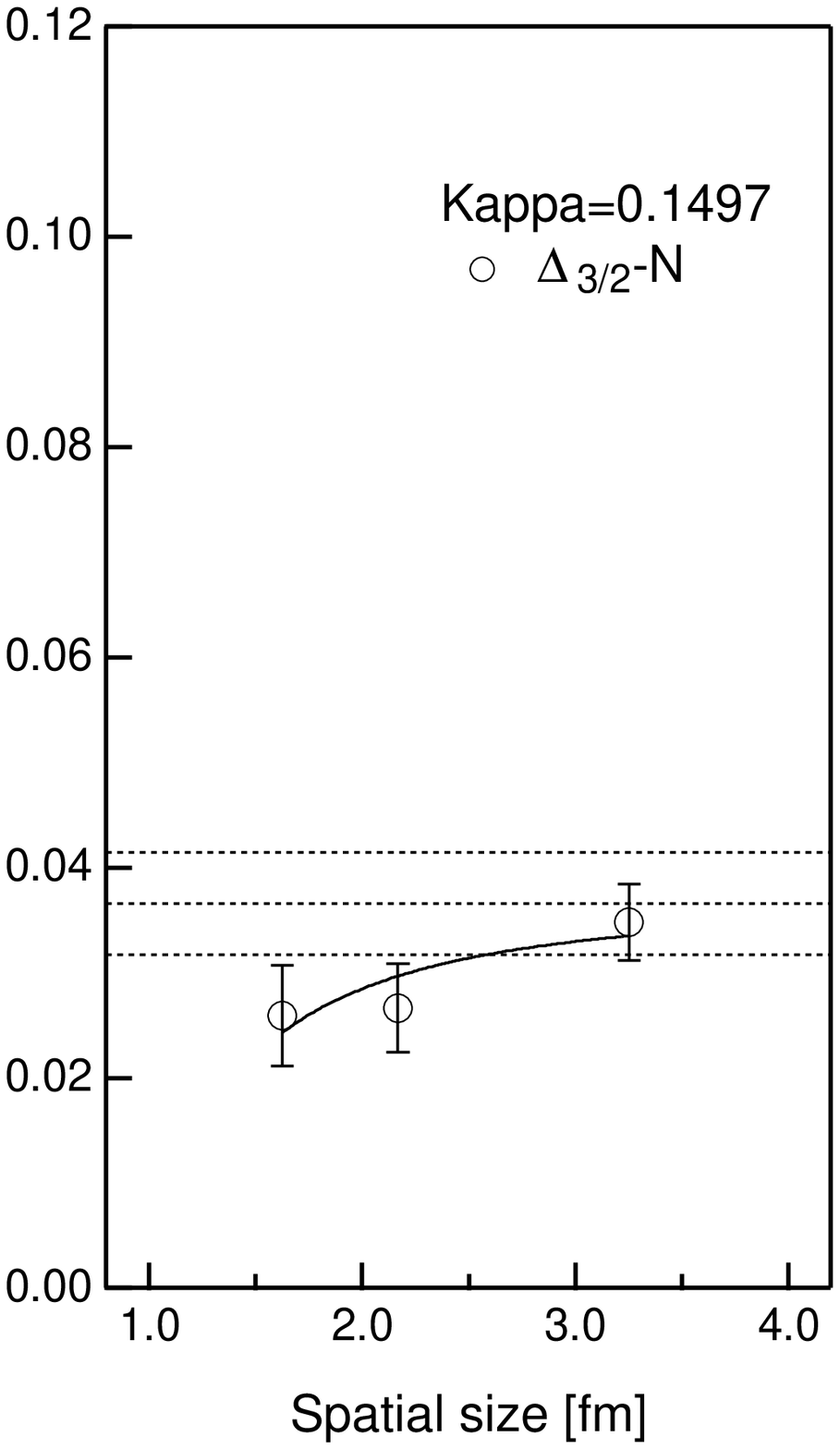}
\linebreak
\vfill
\end{minipage}
\end{center}
\caption{Masses of spin-1/2 $\Delta$ baryons in lattice
unit as functions of spatial lattice size in the physical unit for $\kappa=0.1506$
(left figure) and $\kappa=0.1497$ (right figure). Solid curves
are fits of the form $aM_L=aM_{\infty}+cL^{-n}$ with the value $n=2$.
Horizontal dashed lines represent the mean value of the infinite volume 
extrapolation and one standard deviation from it.
}
\label{fig:InfVol.Hyp}
\end{figure}

\begin{figure}[htbp]
\begin{center}
\begin{minipage}[b]{5.25in}
\includegraphics[width=2.0in]{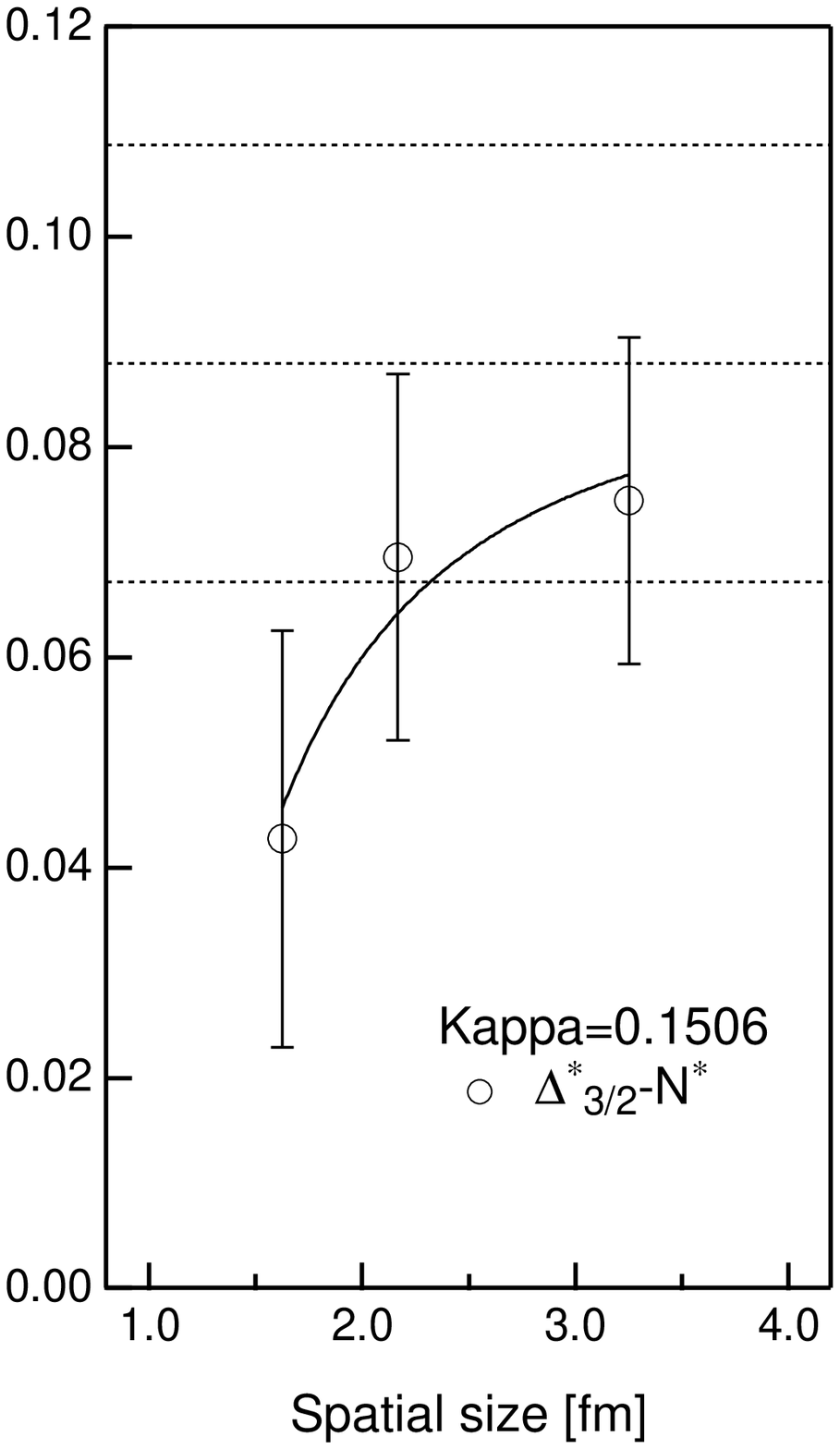}
\includegraphics[width=2.0in]{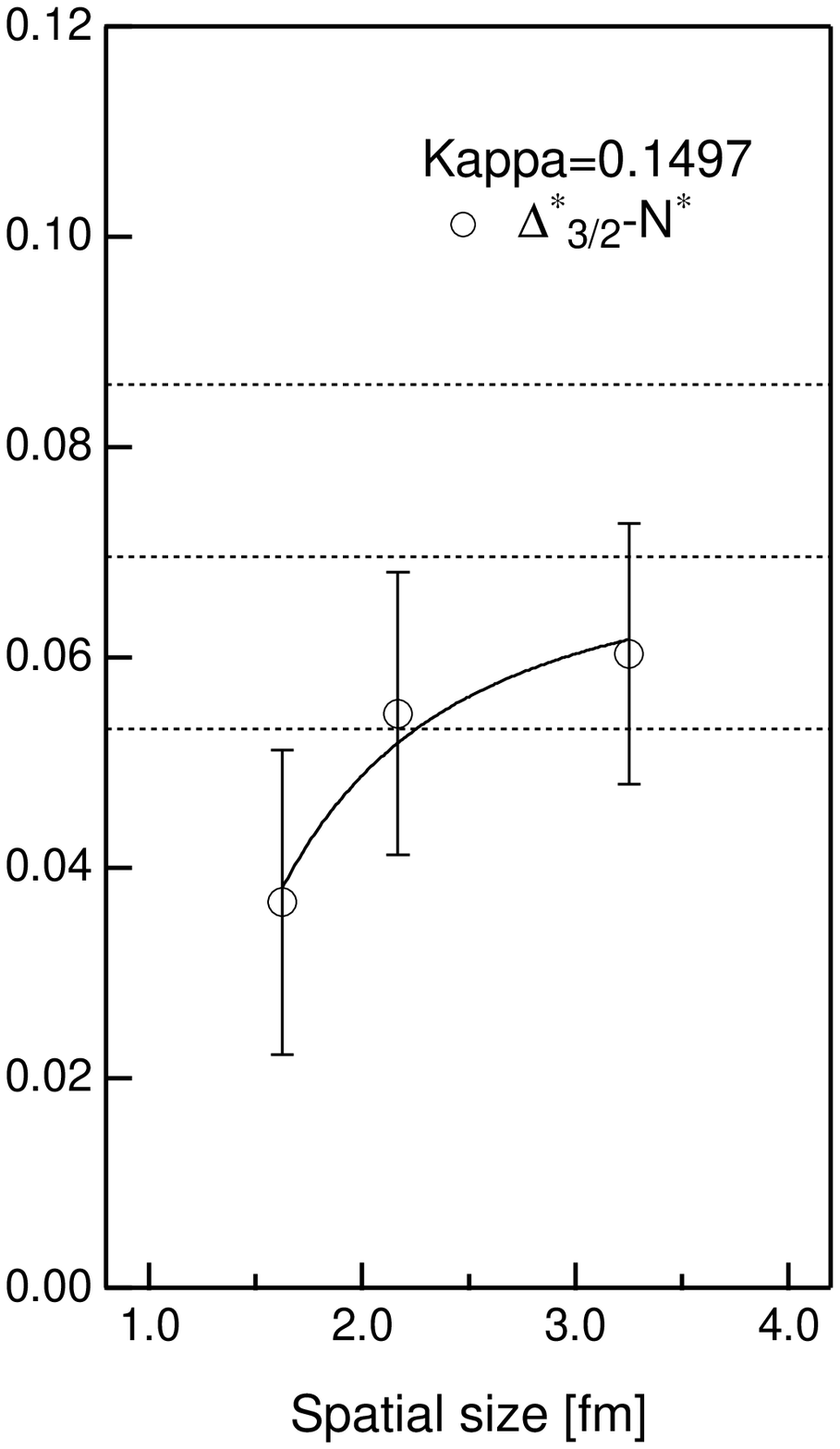}
\linebreak
\vfill
\end{minipage}
\end{center}
\caption{Masses of spin-1/2 $\Delta$ baryons in lattice
unit as functions of spatial lattice size in the physical unit for $\kappa=0.1506$
(left figure) and $\kappa=0.1497$ (right figure). 
Solid curves and dashed lines are defined as in 
Figure~\ref{fig:InfVol.Hyp}.
}
\label{fig:InfVol.HypSTR}
\end{figure}

\begin{figure}[htbp]
\begin{center}
\includegraphics[width=3.0in]{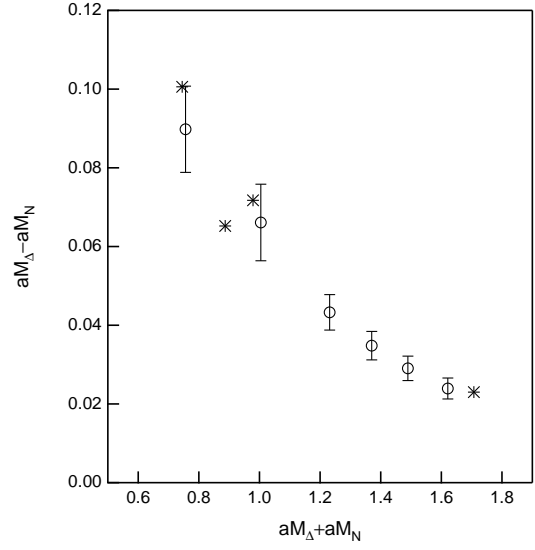}
\end{center}
\caption{Hyperfine splitting between the $N$ and $\Delta_{3/2}$ states
as a function of $M_{\Delta}+M_{N}$. 
All calculations are done for three degenerate valence-quarks.
The experimental values in the wide range from the light (up, down) sector to 
the charm sector are marked by stars.
}
\label{fig:HyperFine}
\end{figure}

\end{document}